\documentclass[aps,pra,superscriptaddress]{revtex4-2}
\pdfoutput=1 %
\usepackage[utf8]{inputenc}
\usepackage{fullpage}
\usepackage{subcaption}
\usepackage[english]{babel}
\usepackage{amsmath}
\usepackage{physics}
\usepackage{amsthm}
\usepackage{amsfonts}
\usepackage{amssymb}
\usepackage{bbold}
\usepackage{mathrsfs}
\usepackage{mathbbol}
\usepackage[toc,page]{appendix}
\usepackage{subcaption}
\usepackage{graphicx}
\usepackage{bbding}
\usepackage{esvect}
\usepackage{titlesec} 
\usepackage{comment}
\usepackage{natbib}
\usepackage{enumitem}
\usepackage{commath}
\usepackage{enumitem}
\usepackage{ulem}
\usepackage{siunitx}
\usepackage{braket}
\usepackage{phaistos}
\usepackage{overpic}
\usepackage{comment}
\usepackage[colorlinks=true
  ,urlcolor=blue
  ,anchorcolor=blue
  ,citecolor=blue
  ,filecolor=blue
  ,linkcolor=blue
  ,menucolor=blue
  ,pagecolor=blue
  ,linktocpage=true
  ,pdfproducer=medialab
  ,pdfa=true
]{hyperref}
\usepackage{xcolor}

\usepackage{tikz}

\usepackage{environ}
\NewEnviron{myequation}{%
\begin{equation*}
\scalebox{1.7}{$\BODY$}
\end{equation*}
}

\usepackage{bbm}

\newcommand{\Hil}{\mathcal{H}}
\newcommand{\HC}{(HC)_{1,2}}
\newcommand{\barHC}{(\overline{HC})_{1,2}}
\newcommand{\Stab}{\textnormal{Stab}}
\def\ket#1{{|{#1}\rangle}}

\begin{document}

\title{Bounding Entanglement Entropy with Clifford Double Cosets}

\author{Cynthia Keeler}
\affiliation{Department of Physics, Arizona State University,
Tempe, AZ 85281, USA}

\author{William Munizzi}
\affiliation{Department of Physics, Arizona State University,
Tempe, AZ 85281, USA}

\author{Jason Pollack}
\affiliation{Department of Electrical Engineering and Computer Science, Syracuse University, NY 13210, USA}

\begin{abstract}
    Following on our previous work \cite{Keeler2022,Keeler:2023xcx} studying the orbits of quantum states under Clifford circuits via `reachability graphs', we introduce `contracted graphs' whose vertices represent classes of quantum states with the same entropy vector.  These contracted graphs represent the double cosets of the Clifford group, where the left cosets are built from the stabilizer subgroup of the starting state and the right cosets are built from the entropy-preserving operators. We study contracted graphs for stabilizer states, as well as W states and Dicke states, discussing how the diameter of a state's contracted graph constrains the `entropic diversity' of its $2$-qubit Clifford orbit. We derive an upper bound on the number of entropy vectors that can be generated using any $n$-qubit Clifford circuit, for any quantum state. We speculate on the holographic implications for the relative proximity of gravitational duals of states within the same Clifford orbit. Although we concentrate on how entropy evolves under the Clifford group, our double-coset formalism, and thus the contracted graph picture, is extendable to generic gate sets and generic state properties. 
\end{abstract}

\maketitle
\flushbottom



\section{Introduction}

One primary goal of quantum computation is to outperform classical computers: that is, for certain tasks, to take a classical input and compute a classical output more rapidly, or efficiently, than any known classical algorithm. (In recent years, this goal has been achieved or brought within reach for certain sets of problems \cite{arute2019quantum,zhong2020quantum}.) Intuitively, quantum computers can only do better on these tasks because they're doing something intrinsically \textit{quantum}: if they weren't, they couldn't outperform the classical method. Formalizing this intuitive result is an object of ongoing research: precisely what feature of a particular quantum algorithm allows it to gain an advantage?
\begin{figure}
  \begin{minipage}{.4\textwidth}
    \centering
    \includegraphics[width=6cm,height=6cm]{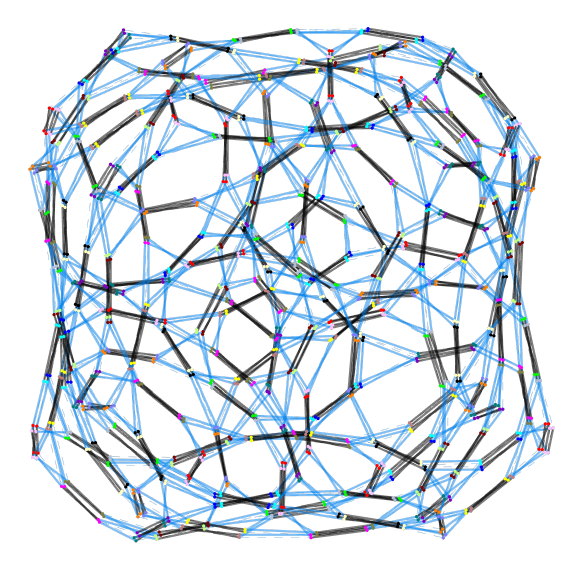}
  \end{minipage}
  \begin{minipage}{.15\textwidth}
    \huge{\begin{myequation}
     {\rightarrow}
    \end{myequation}}
  \end{minipage}
    \begin{minipage}{.4\textwidth}
    \centering
    \includegraphics[width=6cm,height=6cm]{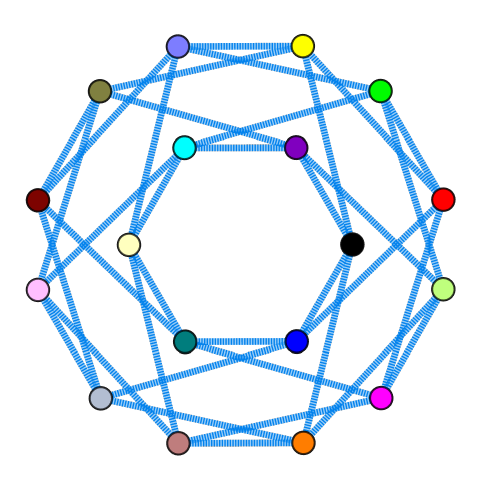}
  \end{minipage}
  \begin{minipage}{.35\textwidth}
    \begin{myequation}
     \quad G/Stab_{G}(\ket{\psi})
    \end{myequation}
  \end{minipage}
  \begin{minipage}{.21\textwidth}
    \begin{myequation}
     \quad
    \end{myequation}
  \end{minipage}
    \begin{minipage}{.35\textwidth}
    \begin{myequation}
     H \backslash G/Stab_{G}(\ket{\psi})
    \end{myequation}
  \end{minipage}
\caption{A reachability graph and its reduction to a contracted graph. In this example, discussed in more detail in Figure \ref{G1152WithContractedGraph}, $G$ is the subgroup of the two-qubit Clifford group generated by Hadamard and $CNOT$ gates and $H$ is the set of operations which leave entropy vectors unchanged.}
\label{TitleFigContractedGraph}
\end{figure}

Setting aside not-even-wrong explanations like ``quantum computers act on each term in a superposition simultaneously," the folk wisdom is that the source of quantum advantage has something to do with interference, superposition, and entanglement. This appealing picture is challenged by the famous result that Clifford circuits, which are generated by the action of one-qubit Hadamard and phase gates and the two-qubit $CNOT$ gate on stabilizer states, can be efficiently classically simulated \cite{Gottesman1998,aaronson2004improved}. That is, even though Clifford circuits can, via $CNOT$ gate applications, produce entanglement, they \textit{can't} give quantum speedups. Evidently, if some kind of entanglement is the key to quantum advantage, the type produced by Clifford gates doesn't suffice.

In order to understand the evolution of entanglement as a state is evolved through a quantum circuit, it's useful to track the \textit{entropy vector}, which characterizes the entanglement entropy of every subsystem of the state. In a recent series of papers, we have investigated how the entropy vector changes under the restricted action of Clifford gates acting on the first two qubits of a state. We first obtained \cite{Keeler2022} the \textit{reachability graphs}, colored by entropy vector, which show how stabilizer states evolve under the action of the two-qubit Clifford group $\mathcal{C}_2$ and its subgroups. In our second paper \cite{Keeler:2023xcx}, having better understood the underlying group-theoretic structures from which the reachability graphs are attained, we were able to find a representation of $\mathcal{C}_2$ as generated by the Clifford gates, as well as explore the reachability graphs produced from initial non-stabilizer states.

Although reachability graphs are useful for directly showing the action of explicit circuits and explicit states, they fail to fully illuminate the paths by which the entropy vector can change. The problem, in short, is that some circuits, even when they contain $CNOT$ gates, fail to change the entropy. For example, one defining relation of $\mathcal{C}_2$ is \cite{Keeler:2023xcx}
\begin{equation}
    \left(CNOT_{1,2}P_2 \right)^4 = P_1^2.
\end{equation}
Hence the structure of reachability graphs by themselves can only loosely bound how the entropy vector might change.

In this paper, we accordingly pass to a more concise graphical representation, the \textit{contracted graphs}, whose vertices represent not single states but classes of states with the same entropy vector. We show how to construct these graphs from the \textit{double cosets} of the Clifford group $\mathcal{C}_2$ and its cosets. An example of this procedure is shown in Figure \ref{TitleFigContractedGraph}. Our protocol for constructing contracted graphs is easily generalized to groups beyond the Clifford group and state properties beyond the entropy vector, and might be of use for other applications.

The remainder of this paper is organized as follows. In Section \ref{sec:review}, we review the Clifford group and stabilizer formalism, as well as the group-theoretic concepts of cosets and double cosets. We also recall the objects used in our previous papers: Cayley graphs, reachability graphs, and entropy vectors. In Section \ref{sec:building}, we give a general procedure for constructing the contracted graphs which retain information about entropy-changing operations in a group. In Section \ref{ContractedGraphsSection}, we apply this procedure to $\mathcal{C}_2$ and its subgroup $\HC$. For each of the reachability graphs in our previous papers, we obtain the resulting contracted graph, and show how these combine together under the action of the full Clifford group. In Section \ref{sec:diversity} we consider the diameter and entropic diversity of the reachability graphs, and discuss implications for the available transformations on a dual geometry via holography. In Section \ref{sec:discussion} we conclude and discuss future work. An appendix collects additional details of our computations.

\section{Review}\label{sec:review}

\subsection{Clifford Group and Stabilizer Formalism}

The Pauli matrices are a set of unitary and Hermitian operators, defined in the computational basis $\{\ket{0},\ket{1}\}$ as
\begin{equation}\label{PauliMatrices}
    \mathbb{1}\equiv\begin{bmatrix}1&0\\0&1\end{bmatrix}, \,\, \sigma_X\equiv\begin{bmatrix}0&1\\1&0\end{bmatrix}, \,\,
    \sigma_Y\equiv\begin{bmatrix}0&-i\\i&0\end{bmatrix}, \,\,
    \sigma_Z\equiv\begin{bmatrix}1&0\\0&-1\end{bmatrix}.
\end{equation}
The multiplicative matrix group generated by $\sigma_X,\,\sigma_Y,$ and $\sigma_Z$ is known as the single-qubit Pauli group $\Pi_1$, which we write
\begin{equation}
    \Pi_1 \equiv \langle \sigma_X,\,\sigma_Y,\,\sigma_Z \rangle.
\end{equation}
When $\Pi_1$ acts on a Hilbert space $\Hil \equiv \mathbb{C}^2$, in the fixed basis spanned by $\{\ket{0},\,\ket{1}\}$, it generates the algebra of all linear operations on $\Hil$.

The Clifford group is likewise a multiplicative matrix group, generated by the Hadamard, phase, and CNOT operations:
\begin{equation}
    H\equiv \frac{1}{\sqrt{2}}\begin{bmatrix}1&1\\1&-1\end{bmatrix}, \quad P\equiv \begin{bmatrix}1&0\\0&i\end{bmatrix}, \quad     C_{i,j} \equiv \begin{bmatrix}
            1 & 0 & 0 & 0\\
            0 & 1 & 0 & 0\\
	    0 & 0 & 0 & 1\\
	    0 & 0 & 1 & 0
            \end{bmatrix}.
\end{equation}
The CNOT gate is a bi-local operation which, depending on the state of one qubit, the control bit, may act with a $\sigma_X$ operation on a second qubit, the target bit. For the gate $C_{i,j}$, the first subscript index denotes the control bit and the second subscript the target bit. We define the single qubit Clifford group $\mathcal{C}_1$ as the group $\langle H,\,P \rangle$. Elements of $\mathcal{C}_1$ act as automorphisms on $\Pi_1$ under conjugation; hence $\mathcal{C}_1$ is contained in the normalizer of $\Pi_1$ in $L(\Hil)$.

When considering the action of the Pauli and Clifford groups on multi-qubit systems, we compose strings of operators which act collectively on an $n$-qubit state. For an element of $\Pi_1$ which acts locally on the $k^{th}$ qubit in an $n$-qubit system, for example, we write
\begin{equation}\label{PauliString}
    I^1\otimes\ldots\otimes I^{k-1} \otimes \sigma_X^k \otimes I^{k+1} \otimes \ldots \otimes I^n.
\end{equation}
Eq. \eqref{PauliString} is referred to as a Pauli string, where the weight of each string counts the number of non-identity insertions. The multiplicative group generated by all Pauli strings of weight $1$ is the $n$-qubit Pauli group $\Pi_n$.

We similarly can extend the action of $\mathcal{C}_1$ to multiple qubits, now incorporating $C_{i,j}$ into the generating set. Composing Clifford strings analogously to Eq. \eqref{PauliString}, we define the $n$-qubit Clifford group $\mathcal{C}_n$ as\footnote{This is not the minimal generating set for the Clifford group, since some Clifford gates can be written in terms of the others, c.f.\ Eqs.\ (4.5, 4.6) of \cite{Keeler:2023xcx}. A more minimal definition of the Clifford group is $C_n\equiv\langle\{H_i,P_1,C_{i,j}\}\rangle$ where $i\in\{1\ldots n\}$, $j>i$.}
\begin{equation}
    \mathcal{C}_n \equiv \langle H_1,...,\,H_n,\,P_1,...,P_n,\,C_{1,2},\,C_{2,1},...,\,C_{n-1,n},\,C_{n,n-1}\rangle.
\end{equation}
When indicating the action of some local gate, Hadamard or phase, the gate subscript denotes which qubit the gate acts on, e.g.\ $H_1$ for the action of Hadamard on the first qubit of an $n$-qubit system.

Beginning with any $n$-qubit computational basis state, e.g.\ $\ket{0}^{\otimes n}$, the group $\mathcal{C}_n$ is sufficient to generate the full set of $n$-qubit stabilizer states. The stabilizer states comprise elements of $\Hil$ which are left invariant under a $2^n$ element subgroup of $\Pi_n$. As we noted in the introduction, stabilizer states are notable in quantum computing as a set of quantum systems which can be efficiently simulated with classical computing \cite{aaronson2004improved,Gottesman1997}. Since the group $\mathcal{C}_n$ is finite, the set of $n$-qubit stabilizer states $S_n$ is also finite \cite{doi:10.1063/1.4818950} and has order given by
\begin{equation}\label{StabilizerSetSize}
    \left|S_n\right| = 2^n \prod_{k=0}^{n-1} (2^{n-k}+1).
\end{equation}

\subsection{Cosets and Double Cosets}

Throughout this paper we support our graph models with parallel group-theoretic arguments. Many of our explanations make substantial use of coset and double coset constructions, which we review here. We also take this opportunity to set notation and establish language that will be used throughout the remainder of the paper.

Let $G$ be a group and $K \leq G$ an arbitrary subgroup. The set of all left cosets of $K$ in $G$ are constructed as
\begin{equation}\label{LeftCosetDefintion}
 g\cdot K, \quad  \forall g \in G.
\end{equation}
Each left coset built in Eq.\ \eqref{LeftCosetDefintion} is an equivalence class of elements $[g_i]$, which are equivalent under $K$ group action on the right:
\begin{equation}\label{LeftCosetEquivalenceRelation}
    g_i \sim g_j \Longleftrightarrow \exists \,k \in K : g_i = g_j k.
\end{equation}
Any two cosets $[g_i]$ in $g\cdot K$ must be either equal or disjoint, and every $g \in G$ must be found in exactly one equivalence class. As a result, the set of all $[g_i]$ gives a partition of $G$.

Eqs.\ \eqref{LeftCosetDefintion} and \eqref{LeftCosetEquivalenceRelation}, as well as the accompanying explanations, apply analogously when generating all right cosets $H \cdot g$, for arbitrary $H \leq G$. We build all right cosets by computing $H \cdot g$, for every $g \in G$, where each equivalence class $[g_i]$ is now determined by left subgroup action
\begin{equation}\label{RightCosetEquivalenceRelation}
    g_i \sim g_j \Longleftrightarrow \exists \,h \in H : g_i = hg_j.
\end{equation}
When $H \leq G$ is normal in $G$, the left and right cosets%
\footnote{We use the notation $G/H$ to indicate the set of left cosets of $H$ in $G$, as described in Eq. \eqref{LeftCosetDefintion}, and $H \backslash G$ to indicate the set of right cosets of $H$ in $G$, as in Eq. \eqref{RightCosetEquivalenceRelation}.} %
are equal, and $G/H$ forms a group under the same binary operation which defines $G$.

Two subgroups $H,K \leq G$ can be used to construct double cosets of $G$. We build each $(H,K)$ double coset by acting on $g \in G$ on the right by subgroup $K$, and on the left by $H$, explicitly
\begin{equation}\label{DoubleCosetDefinition}
 H\cdot g \cdot K, \quad  \forall g \in G.
\end{equation}
The double coset space built using Eq.\ \eqref{DoubleCosetDefinition} is denoted $H \backslash G / K$, and is described by the equivalence relation
\begin{equation}\label{EqClassDoubleCoset}
    g_i \sim g_j \Longleftrightarrow \exists \,h \in H,\, k \in K : g_i = hg_jk.
\end{equation}

In order to utilize the above coset constructions in this paper, we invoke several foundational group theory concepts (see e.g.\ \cite{Alperin1995}). First, for a finite group $G$, the order of any subgroup $K \leq G$ divides the order of $G$ by Lagrange's theorem
\begin{equation}\label{LagrangeTheorem}
    \frac{|G|}{|K|} = [G:K], \quad \forall K\leq G,
\end{equation}
where $[G:K] \in \mathbb{N}$ is the number of left (or right) cosets of $K$ in $G$. When acting with $G$ on a set $X$, the orbit-stabilizer theorem fixes the size of each orbit $[G \cdot x]$ to be
\begin{equation}\label{OrbitStabilizerTheorem}
    [G \cdot x] = [G:K_x] = \frac{|G|}{|K_x|}, \quad \forall x\leq X,
\end{equation}
where $K_x \leq G$ is the set of elements which map $x \in X$ to itself.

We can likewise use Eq.\ \eqref{LagrangeTheorem} with Eq.\ \eqref{OrbitStabilizerTheorem} to compute the order of a double coset space, i.e.\ the orbit of all left (or right) cosets under left (or right) subgroup action. For finite $G$ and subgroups $H,K \leq G$, the order%
\footnote{Note that a direct application of Lagrange's theorem to the order of a double coset space is false, i.e.\ the order of a double coset space of $G$ does not necessarily divide $|G|$.} %
of $H \backslash G / K$ is computed as
\begin{equation}\label{DoubleCosetOrder}
     |H \backslash G / K| = \frac{1}{|H||K|} \sum_{(h,k) \in H \times K} |G^{(h,k)}|, 
\end{equation}
where $G^{(h,k)} = hG^{\mathbb{1}}k$, and $G^{\mathbb{1}}$ is a set consisting of a representative element from each equivalence class under Eq.\ \eqref{EqClassDoubleCoset}. The sum in Eq.\ \eqref{DoubleCosetOrder} is taken over all ordered pairs $(h,k)$ of $h \in H$ and $k \in K$, and importantly the order of each $G^{(h,k)}$ can be different from one another.

\subsection{Cayley Graphs and Reachability Graphs}\label{CayleyGraphReview}

A \textit{Cayley graph} encodes in graphical form the structure of a group. For a group $G$ and a chosen set of generators, we construct the Cayley graph of $G$ by assigning a vertex for every $g \in G$, and an edge%
\footnote{Formally, each edge in a Cayley graph is directed. However, for improved legibility, we will often represent group generators which are their own inverse using undirected edges.} %
leaving each vertex for every generator of $G$. When $G$ corresponds to a set of quantum operators acting on a Hilbert space, paths in the Cayley graph represent quantum circuits that can be composed using the generating gate set. Different paths which start and end on the same pair of vertices indicate sequences of operators whose action on every quantum state is identical. Loops in a Cayley graph represent operations equivalent to the identity.

For a group $G \subset L(\Hil)$, we define the stabilizer subgroup $\Stab_G(\ket{\psi})$ of some $\ket{\psi} \in \Hil$ as the subset of elements $g \in G$ which leave $\ket{\psi}$ unchanged,
\begin{equation}
    \Stab_G(\ket{\psi}) \equiv \{g \in G \,|\, g\ket{\psi} = \psi \}.
\end{equation}
In other words, the subgroup $\Stab_G(\ket{\psi})$ consists of all $g \in G$ for which $\ket{\psi}$ is an eigenvector with eigenvalue $+1$.

Reachability graphs can be obtained more generally as quotients of Cayley graphs \cite{Keeler:2023xcx,githubStab,githubCayley}. To perform this procedure, we first identify a group $G \in L(\Hil)$ to act on a Hilbert space $\Hil$, and a generating set for $G$. We then first quotient $G$ by the subgroup of elements which act as an overall phase on the Hilbert space. For $\mathcal{C}_n$, this is the subgroup $\langle \omega \rangle$, where
\begin{equation}\label{OmegaQuotient}
    \omega \equiv \left(H_iP_i\right)^3 = e^{i\pi/4}\mathbb{1}.
\end{equation}
Here, $\mathbb{1}$ refers to the identity element of $G$, whose matrix representation is the identity matrix. Once we have removed overall phase and constructed the quotient group%
\footnote{Since $\langle \omega \rangle < \mathcal{C}_n$ is normal (and in fact in the center of $G$), modding by $\langle \omega \rangle$ builds a proper quotient. It therefore does not matter whether we apply $\langle \omega \rangle$ on the left or right of $G$ when building cosets, nor does it affect any subsequent double coset construction.} %
$\bar{G} = G/\langle \omega \rangle$, we identify a state $\ket{\psi} \in \Hil$. Selecting  $\ket{\psi}$ immediately defines the stabilizer subgroup $\Stab_{\bar{G}}(\ket{\psi})$. We then construct the left coset space $\bar{G}/\Stab_{\bar{G}}(\ket{\psi})$ whose elements are
\begin{equation}\label{StabGroupQuotient}
    g\cdot\Stab_{\bar{G}}(\ket{\psi}) \quad \forall g \in \bar{G}.
\end{equation}

To graphically represent this procedure, we begin by selecting a group $\bar{G}$. We construct a graph $\Gamma$, with vertex set $V$, such that each $v \in V$ corresponds to an element of $\bar{G}$. We then quotient $\Gamma$ by gluing together vertices $u,\,v \in V$, if $u$ and $v$ represent elements of $\bar{G}$ that share the same coset under, for example, Eq.\ \eqref{StabGroupQuotient}. In this way, vertices of the quotient graph represent cosets of $\bar{G}$.


While the graphs in this paper often represent groups, constructing a graph quotient is not equivalent to quotienting a group. Building a group quotient requires modding by a normal subgroup, which ensures that the left and right coset spaces of the chosen subgroup are equal, preserving the original group action in the quotient group. We do not impose such a requirement when building graph quotients in this paper, even when our graphs illustrate the relation between groups of operators. We distinguish graph quotients from group quotients wherever potential confusion could occur.

\subsection{Entropy Vectors and Entropy Cones}

For a state $\ket{\psi} \in \Hil$, and some specified factorization for $\Hil$, we can compute the von Neumann entropy of the associated density matrix:
\begin{equation}\label{vonNeumannEntropy}
   S_{\psi} \equiv -\Tr \left( \rho_{\psi} \log \left(\rho_{\psi}\right) \right),
\end{equation}
where $\rho_{\psi} \equiv \ket{\psi}\bra{\psi}$. For $\ket{\psi}$ a pure state, the property $\rho_{\psi}^2 = \rho_{\psi}$ implies $S_{\psi} = 0$. Throughout this paper, we measure information in \textit{bits}, and entropies in Eq.\ \eqref{vonNeumannEntropy} are computed with $\log_2$.

For a multi-partite pure state $\ket{\psi}$, we can observe non-zero entanglement entropy among complementary subsystems of $\ket{\psi}$. Let $\ket{\psi}$ be some $n$-party pure state, and let $I$ denote an $\ell$-party subsystem of $\ket{\psi}$. We can compute the entanglement entropy between $I$ and its $(n-\ell)$-party complement, $\bar{I}$, using
\begin{equation}\label{SubsystemEntropyDefinition}
   S_{I} = -\Tr \left(\rho_{I} \log \left(\rho_{I}\right) \right).
\end{equation}
The object $\rho_{I}$ in Eq.\ \eqref{SubsystemEntropyDefinition} indicates the reduced density matrix of subsystem $I$, which is computed by tracing out the complement subsystem $\bar{I}$.

In general, there are $2^n-1$ possible subsystem entropies we can compute for any $n$-qubit pure state $\ket{\psi}$. Computing each $S_{I}$, using Eq.\ \eqref{SubsystemEntropyDefinition}, and arranging all entropies into an ordered tuple defines the entropy vector $\Vec{S}\left(\ket{\psi} \right)$. As an example, consider the $4$-qubit pure state $\ket{\psi}$, where $\Vec{S}\left(\ket{\psi} \right)$ is defined
\small{
\begin{equation}\label{EntropyVectorExample}
    \Vec{S} = (S_A,S_B,S_C,S_O; S_{AB},S_{AC},S_{AO},S_{BC},S_{BO},S_{CO};S_{ABC},S_{ABO},S_{ACO},S_{BCO}; S_{ABCO}),
\end{equation}}\normalsize
where again each component is computed using Eq.\ \eqref{SubsystemEntropyDefinition}. In Eq.\ \eqref{EntropyVectorExample} we use a semicolon to separate entropy components for subregions of distinct cardinality $|I|$. Additionally, for an $n$-qubit state it is customary to denote the $n^{th}$ subsystem using $O$, as this region acts as a purifier for the other $n-1$ parties. 

For an $n$-party system, each entropy vector contains $2^n-1$ components, with the first $n$ components representing single-qubit subsystems. We list entropy vector components in lexicographic order: with the first region denoted $A$, the second region denoted $B$, and so forth. Unlike what is sometimes found in the literature, we use $O$ to represent a smaller bipartition, instead of the one which does not contain the purifier. For example, in Eq.\ \eqref{EntropyVectorExample} we declare $O$ a single-party subsystem which purifies $ABC$, and write $S_O$ in place of $S_{ABC}$ among the single-party entries of the entropy vector.

When $\ket{\psi}$ is a pure state, the condition $S_{\psi} = 0$ implies an additional equivalence between entropies of complement subsystems
\begin{equation}\label{SubregionComplement}
    S_{I} = S_{\bar{I}}.
\end{equation}
Using Eq.\ \eqref{SubregionComplement} we can write $\Vec{S}\left(\ket{\psi} \right)$, for a pure state $\ket{\psi}$, using only $2^{n-1}-1$ entropies. For example, the entropy vector in Eq.\ \eqref{EntropyVectorExample} simplifies to the form
\begin{equation}\label{ReducedEntropyVectorExample}
    \Vec{S} = (S_A,S_B,S_C,S_O; S_{AB},S_{AC},S_{AO}).
\end{equation}
Since we are always considering pure states in this paper, all entropy vectors are written using the reduced notation in Eq.\ \eqref{ReducedEntropyVectorExample}.

\section{Building Contracted Graphs}\label{sec:building}

We now define a procedure to quotient reachability graphs by operations which preserve some specified property of a quantum system. In this paper we focus on the evolution of entanglement entropy under the action of the Clifford group; however, this prescription is sufficiently general to study any state property%
\footnote{In this work, the term \textit{state property} refers to anything computable from knowledge of the state, along with some additional information such as a specified factorization of the Hilbert space. We do not restrict analysis to properties which are observables; in fact, the main property discussed in this paper, the entropy vector, is not itself an observable.\label{fn:state_property}} %
under the action of any finitely-generated group.

We build a \textit{contracted graph} by identifying vertices in a reachability graph which are connected by entropy-preserving circuits. In this way, a contracted graph details the evolution of a state's entropy vector under the chosen gate set. It is important to note that group elements do not act directly on entropy vectors themselves, but act on the underlying quantum states, whose transformation subsequently affects the associated entropy vectors. The number of vertices in a contracted graph gives a strict upper bound on the number of different entanglement vector values reachable via circuits constructed using the chosen gate set. We will later use contracted graphs to derive an upper bound on entropy vector variation in Clifford circuits.

We now give an algorithm for generating contracted graphs. 

\begin{enumerate}
    \item We first select a group $G$, and a generating set for $G$, as well as a property of our quantum system we wish to study under the action of $G$.

    \item We quotient $G$ by the subgroup which acts as a global phase on the group, such as in Eq.\ \eqref{OmegaQuotient}. We next build the Cayley graph for $\bar{G}$ by assigning a vertex for every $g \in \bar{G}$, and a directed edge for each generator action on an element $g \in \bar{G}$.
    
    \item Next, we construct the reachability graph for some $\ket{\psi}$ under the action of $\bar{G}$, as detailed in Subsection \ref{CayleyGraphReview}, which we denote%
    \footnote{A more precise notation for such reachability graphs would be $\mathcal{R}\left(\textnormal{Stab}_{\bar{G}}\left(\ket{\psi}\right)\right)$, however we choose $\mathcal{R}_{\bar{G}}\left(\ket{\psi}\right)$ instead for brevity.} %
    $\mathcal{R}_{\bar{G}}\left(\ket{\psi}\right)$.
    We determine the stabilizer subgroup $\Stab_{\bar{G}}\left(\ket{\psi}\right)$ for $\ket{\psi}$, and generate the left coset space $\bar{G}/ \Stab_{\bar{G}}\left( \ket{\psi} \right)$ using the equivalence relation
    \begin{equation}\label{StabilizerEquivalence}
        g_i \sim g_j \Longleftrightarrow \exists \,s \in \Stab_{\bar{G}}\left(\ket{\psi}\right) : g_i = g_js.
    \end{equation}

    We glue together vertices in the Cayley graph of $\bar{G}$ that correspond to elements which share an equivalence class $[g_i]$ in $\bar{G}/ \Stab_{\bar{G}}\left( \ket{\psi} \right)$. This graph quotient yields $\mathcal{R}_{\bar{G}}\left(\ket{\psi}\right)$.
    
    \item We now identify the subgroup $H \leq \bar{G}$ of elements that leave the entropy vector of every state invariant. The subgroup $H$ defines the equivalence relation
    \begin{equation}\label{EntropyEquivalence}
        g_i \sim g_j \Longleftrightarrow \exists \,h \in H : g_i = hg_j.
    \end{equation}
    $H$ will contain any element of $\bar{G}$ arising from a generator of ${G}$ which acts as a local gate on a single qubit, since local action cannot modify entanglement.
    
    \item Finally, we build all double cosets $H \backslash \bar{G} / \Stab_{\bar{G}}\left( \ket{\psi} \right)$. We identify all vertices in $\mathcal{R}_{\bar{G}}\left(\ket{\psi}\right)$ which share an equivalence class in $H \backslash \bar{G} / \Stab_{\bar{G}}\left( \ket{\psi} \right)$, and subsequently quotient $\mathcal{R}_{\bar{G}}\left(\ket{\psi}\right)$ to give the final contracted graph.
\end{enumerate}

We generate reachability graphs by building left cosets $\bar{G}/ \Stab_{\bar{G}}\left( \ket{\psi} \right)$, defined by an equivalence up to right subgroup action by $\Stab_{\bar{G}}\left( \ket{\psi} \right)$ as in Eq.\ \eqref{StabilizerEquivalence}. Since $\Stab_{\bar{G}}\left( \ket{\psi} \right)$ acts trivially on $\ket{\psi}$, appending any $s \in \Stab_{\bar{G}}\left( \ket{\psi} \right)$ to the right of any  $g \in \bar{G}$ does not change how $g$ transforms the state $\ket{\psi}$. Conversely, we build a contracted graph by generating right cosets $\bar{G} \backslash H$, with equivalence defined up to left subgroup action as shown in Eq.\ \eqref{EntropyEquivalence}. Every element of $H$ preserves a state's entropy vector, therefore acting on the left of $g\ket{\psi}$ by any $h \in H$ does not change the measurement of the full state entropy vector, for every $g \in \bar{G}$.

Recall that there are two interpretations%
\footnote{Consider two generic states stabilized by only $\mathbb{1} \in \bar{G}$. These two states will have isomorphic $\mathcal{R}_{\bar{G}}\left(\ket{\psi}\right)$, considered identical in the group interpretation. However, these graphs are distinct in the state-orbit interpretation, since here the vertices have specific meaning as states in each orbit.} %
of a reachability graph. By identifying a state $\ket{\psi}$ and group $G$ of operators acting on that state, $\mathcal{R}_{\bar{G}}\left(\ket{\psi}\right)$ represents the orbit of $\ket{\psi}$ under the action of $\bar{G}$. In this state-orbit interpretation, vertices of $\mathcal{R}_{\bar{G}}\left(\ket{\psi}\right)$ represent states reached in the orbit of $\ket{\psi}$. For simplicity, we choose this state-orbit interpretation in this explanatory section. A more general interpretation of reachability graphs exists which defines $\mathcal{R}_G\left(\ket{\psi}\right)$ as a left coset space of the Cayley graph of the abstract group $G$. In this interpretation, vertices represent equivalence classes of $g \in G$ defined by the left coset $g \cdot \Stab_G\left(\ket{\psi}\right)$.

\paragraph{Example:}
For clarity, we now work through an explicit example. Consider the subgroup of the two-qubit Clifford group%
\footnote{For additional detail on this Clifford subgroup see Section 4.2 of \cite{Keeler:2023xcx}, where all group elements are derived using two-qubit Clifford group relations.} %
generated by the $P_2$ and $C_{1,2}$ gates,
\begin{equation}
    G \equiv \langle P_2,\, C_{1,2}\rangle.
\end{equation}
The group $\langle P_2,\, C_{1,2}\rangle$ consists of $32$ elements, specifically
\begin{equation}
   \langle P_2,\, C_{1,2}\rangle = \{P_2^\alpha,\,P_2^\alpha C_{1,2}P_2^\beta,\,C_{1,2}P_2^\gamma C_{1,2}P_2^\alpha\},
\end{equation}
where $\alpha,\, \beta \in \{0,1,2,3\}$ and $\gamma \in \{1,2,3\}$.

\begin{figure}[h]
        \centering
        \includegraphics[width=13cm]{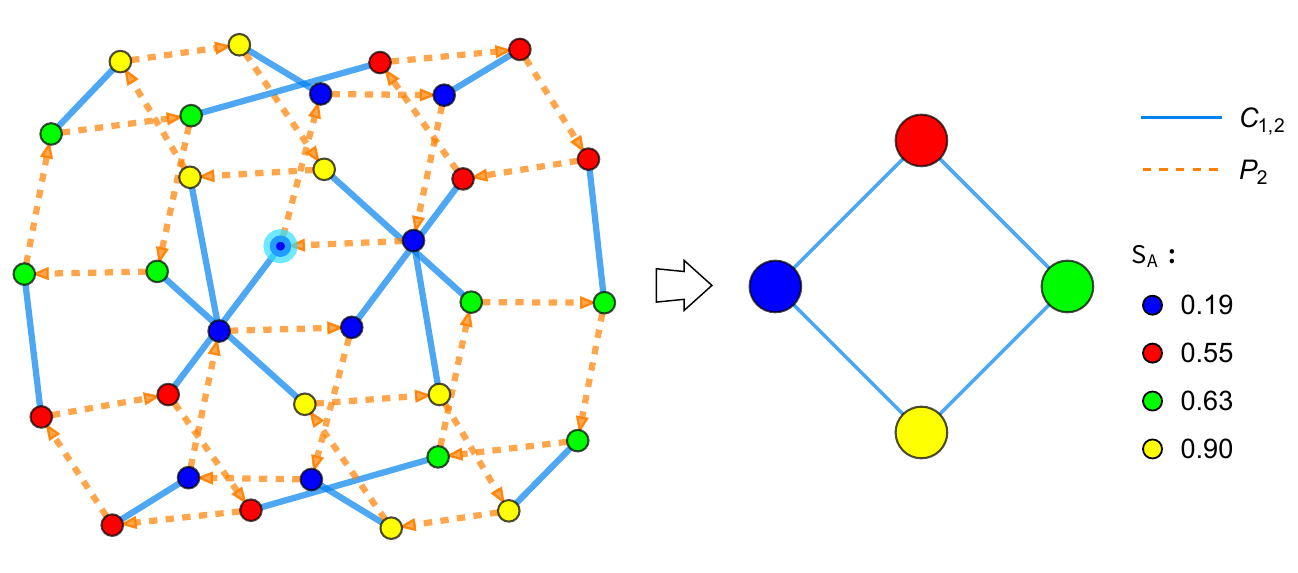}
        \caption{Reachability graph (left) of $\ket{\psi} = \left(\ket{00} + 2\ket{01} + 4\ket{10}+ 3\ket{11}\right)/\sqrt{30}$, highlighted in cyan, under action of $\langle P_2,\ CNOT_{1,2}\rangle$, and its associated contracted graph (right). The contracted graph has $4$ vertices and $4$ edges connecting any two vertices, indicating the entropy vector can maximally change $4$ times under any circuit built of $P_2$ and $CNOT_{1,2}$. The $4$ entropy vector possibilities, defined by Eq.\ \eqref{EntropyVectorExample}, are given in the legend.}
        \label{P2C12CayleyWithContractedGraph}
    \end{figure}

We select the state $\ket{\psi} = \left(\ket{00} + 2\ket{01} + 4\ket{10}+ 3\ket{11}\right)/\sqrt{30}$, which we choose for its particular entropic properties that we discuss in footnote 10. We construct the reachability graph $\mathcal{R}_G\left(\ket{\psi}\right)$ for $\ket{\psi}$, shown in the left panel of Figure \ref{P2C12CayleyWithContractedGraph}. The only element of $G$ which leaves $\ket{\psi}$ invariant is $\mathbb{1}$ in $G$, therefore
\begin{equation}\label{IdentityStabGroup}
    \Stab_G(\ket{\psi}) = \{ \mathbb{1}\}. 
\end{equation}
Since the stabilizer group in Eq.\ \eqref{IdentityStabGroup} consists of just the identity, and is therefore a normal subgroup, the group $\Stab_G(\ket{\psi})$ quotients $G$ and the reachability graph $\mathcal{R}_G\left(\ket{\psi}\right)$ is exactly the $32$-vertex Cayley graph. In the more general case, $\mathcal{R}_G\left(\ket{\psi}\right)$ would not necessarily represent a group quotient, but would represent a left coset space.

We construct the contracted graph of $\mathcal{R}_G\left(\ket{\psi}\right)$ by identifying the elements of $G$ which cannot modify the entropy vector of $\ket{\psi}$. Since the gate $P_2$ acts locally on a single qubit, it can never modify entanglement. Accordingly, we initially contract $\mathcal{R}_G\left(\ket{\psi}\right)$ by gluing together all vertices connected by a $P_2$ edge, represented by the orange dashed lines. Additionally, as we recognized in \cite{Keeler:2023xcx},
\begin{equation}
    \left(C_{1,2}P_2 \right)^4 = P_1^2.
\end{equation}
 Hence all vertices connected by the circuit $\left(C_{1,2}P_2 \right)^4$ must be identified together as well, since $P_1$ likewise does not change a state's entropy vector. The right panel of Figure \ref{P2C12CayleyWithContractedGraph} shows the final contracted graph of $\mathcal{R}_G\left(\ket{\psi}\right)$, which contains $4$ vertices. In this particular example, the contracted graph represents the right coset space of the quotient group $G/\Stab_G(\ket{\psi})$, since $G=\overline{G}$. In general, however, the contracted graph will represent the double coset space $H \backslash G/\Stab_G(\ket{\psi})$, where $G/\Stab_G(\ket{\psi})$ need not be a quotient group.

It is important to note that edges in a contracted graph do not represent any one particular $C_{i,j}$ operation. Instead, every edge bearing a CNOT coloration represents sequences of operations which, at least, include a $C_{i,j}$ gate and are capable of modifying the entropy vector of a sufficiently-general state. In this way, the edges of a contracted graph bound the number of times the entropy vector of a system can change. Since the process of building a contracted graph removes all group elements which leave entanglement entropy unchanged, we are left with a graph structure that represents the orbit of an entropy vector under the group action.

The number of vertices in a contracted graph give an upper bound on the number of distinct entropy vectors which can be generated in a particular reachability graph. For example, the contracted graph in Figure \ref{P2C12CayleyWithContractedGraph} contains $4$ vertices, indicating the maximum number of entropy vectors that can be achieved by acting on $\ket{\psi}$ with $\langle P_2,\, CNOT_{1,2}\rangle$.  The number of vertices in a contracted graph is fixed by the overall group structure of $G$, as well as the group structure of $\Stab_G$; however, the different ways in which those vertices can be colored according to entanglement structure is set by the choice%
\footnote{The fact that this contracted graph contains a unique entropy vector for each of its $4$ vertices, i.e.\ the contracted graph is maximally-colored, is the reason we chose $\ket{\psi}$ as we did.}\label{StateChoiceFootnote} %
of state. While the number of vertices in a contracted graph gives an upper bound on entropic diversity in reachability graphs, there can be multiple entropic colorings of the same graph, depending on factors such as qubit number or the specific state. 

We have defined a procedure for building contracted graphs from the reachability graph of arbitrary state $\ket{\psi}$. When considering a group $G$ which acts on a Hilbert space, we build the reachability graph of $\ket{\psi}$ by decomposing $G$ into left cosets $G/\Stab_G(\ket{\psi})$, with elements equivalent up to action by $\Stab_G(\ket{\psi})$. We build the contracted graph of the $\ket{\psi}$ reachability graph by building the double coset space $H \backslash G /\Stab_G(\ket{\psi})$, for a subgroup $H \leq G$ of elements which preserve a state's entropy vector. 

We have demonstrated how contracted graphs illustrate the evolution of entanglement entropy under the action of some quantum gate set. The number of vertices in a contracted graph gives an upper bound on the maximal number of times an entropy vector can change under the chosen set of gates. We have chosen in this paper to construct contracted graphs from reachability graphs in order to analyze the evolution of state entropy vectors; however, the contraction procedure can be applied directly to Cayley graphs as well. 

In the next section we use the techniques defined above to build contracted graphs for all stabilizer state reachability graphs studied in \cite{Keeler2022,Keeler:2023xcx}, establishing upper bounds on the variation of entanglement entropy in stabilizer state systems. We also extend our analysis beyond stabilizer states, deriving upper bounds on the evolution of entanglement entropy for any quantum state under the action of the Clifford group.

\section{Contracted Clifford Reachability Graphs}\label{ContractedGraphsSection}

In this section, we build contracted graphs to illustrate entropy vector evolution in stabilizer and non-stabilizer state reachability graphs. We begin by first considering stabilizer state reachability graphs under the action of the $\mathcal{C}_2$ subgroup $(\overline{HC})_{1,2} \equiv \overline{\langle H_1,\,H_2,\,C_{1,2},\,C_{2,1} \rangle}$, as studied in \cite{Keeler2022,Keeler:2023xcx}. We demonstrate how the contracted version of each $(\overline{HC})_{1,2}$ reachability graph explains the bounds on entanglement variation observed in our earlier work \cite{Keeler2022}. We then extend our analysis to consider the full action of $\mathcal{C}_2$ on stabilizer states, showing how $\mathcal{C}_2$ contracted graphs constrain the evolution of entanglement entropy in stabilizer systems under any $2$-qubit Clifford circuit. 

We extend our study beyond the stabilizer states to the set of $n$-qubit Dicke states, a class of non-stabilizer quantum states possessing non-trivial stabilizer group under Clifford action \cite{Munizzi:2023ihc}. We construct $(HC)_{1,2}$ and $\mathcal{C}_2$ reachability and contracted graphs for all Dicke states, establishing constraints on entropy vector evolution for such states. Finally we move toward complete generality, deriving an upper bound for the number of entropy vectors that can be realized by any $n$-qubit Clifford circuit, acting on an arbitrary quantum state.

\subsection{Contracted Graphs of $g_{24}$ and $g_{36}$}

The complete set of $n$-qubit stabilizer states can be generated by acting with $\overline{\mathcal{C}}_n$ on the state $\ket{0}^{\otimes n}$. However, since we are motivated to better understand the evolution of entropy vectors in stabilizer systems, we restrict analysis to $\mathcal{C}_2$ and its subgroups, since all entanglement modification in Clifford circuits occurs through bi-local operations. Acting with $\overline{\mathcal{C}}_2$ on $\ket{0}^{\otimes n}$, for $n >1$, generates an orbit of $60$ states.

First, we consider the class of states with stabilizer subgroup%
\footnote{A comprehensive derivation of all stabilizer subgroups, for stabilizer states under the action of $(\overline{HC})_{1,2}$, is given in Section 5.3 of \cite{Keeler:2023xcx}.} %
isomorphic to $\mathcal{S}_{\overline{HC}}(\ket{0}^{\otimes n})\equiv\Stab_{(\overline{HC})_{1,2}}(\ket{0}^{\otimes n})$, under the action of $(\overline{HC})_{1,2}$. The state $\ket{0}^{\otimes n}$, and any other state with stabilizer group isomorphic to $\mathcal{S}_{\overline{HC}}(\ket{0}^{\otimes n})$, has an orbit of $24$ states under $(\overline{HC})_{1,2}$.

\subsubsection{$\left(HC\right)_{1,2}$ Contracted Graphs of $g_{24}$ and $g_{36}$}

\begin{figure}[h]
        \centering
        \includegraphics[width=15cm]{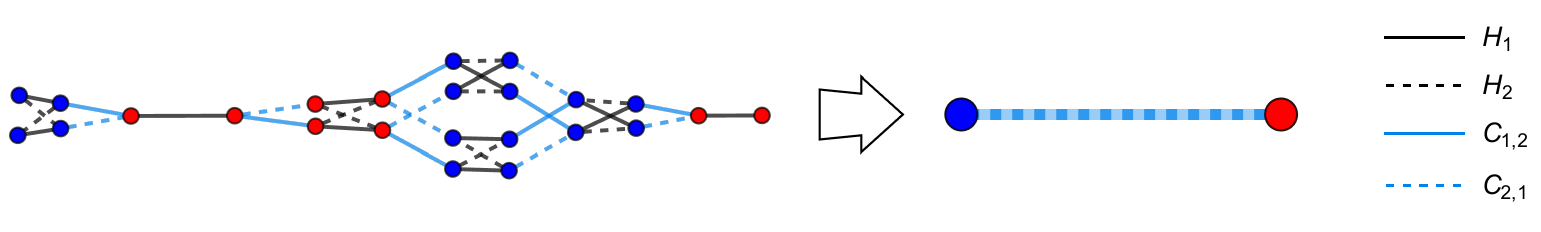}
        \caption{Reachability graph $g_{24}$ (left) and its contracted graph (right). Any state with stabilizer group isomorphic to $\mathcal{S}_{\overline{HC}}(\ket{0}^{\otimes n})$ will have $\HC$ reachability graph $g_{24}$. The $g_{24}$ contracted graph has $2$ vertices, indicating the maximum number of unique entropy vectors that can exist in any $g_{24}$ graph. Each edge in the contracted graph represents a set of entanglement-modifying circuits, each containing at least one CNOT gate.}
        \label{G24WithContractedGraph}
    \end{figure}

The stabilizer subgroup $\mathcal{S}_{\overline{HC}}(\ket{0}^{\otimes n})$ contains $48$ elements. As a result, generating all left cosets of the $1152$-element group $(\overline{HC})_{1,2}$ by $\mathcal{S}_{\overline{HC}}(\ket{0}^{\otimes n})$ builds a coset space of $1152/48 = 24$ equivalence classes. The corresponding $\HC$ reachability graph of $\ket{0}^{\otimes n}$ contains $24$ vertices, which we appropriately term $g_{24}$. The left panel of Figure \ref{G24WithContractedGraph} shows the graph $g_{24}$, which is shared by all states with stabilizer group isomorphic to $\mathcal{S}_{\overline{HC}}(\ket{0}^{\otimes n})$.  

To build the associated contracted graph we quotient $g_{24}$ by all elements of $(\overline{HC})_{1,2}$ which do not modify the entropy vector. One immediate $(\overline{HC})_{1,2}$ subgroup which cannot modify entanglement entropy is $\overline{\langle H_1,\,H_2 \rangle}$, which describes all circuits composed of Hadamard gates acting on two qubits. Additionally, as we recognized in \cite{Keeler:2023xcx}, the relation
\begin{equation}\label{HCLocalAction}
    \left(C_{i,j}H_j \right)^4 = P_i^2,
\end{equation}
demonstrates that certain sequences of Hadamard and CNOT gates are actually equivalent to phase operations. We therefore need to also identify all vertices connected by the circuits in Eq.\ \eqref{HCLocalAction}, since phase operations cannot change entanglement.

After identifying all vertices connected by entropy-preserving edges, the reachability graph $g_{24}$ contracts to a graph with $2$ vertices, shown on the right of Figure \ref{G24WithContractedGraph}. These $2$ vertices represent the $2$ possible entropy vectors that can be reached by all circuits in any $g_{24}$ graph, regardless of qubit number. All states represented by blue vertices in $g_{24}$ are connected by some circuit composed of $H_1,\,H_2,\,P_1^2,$ and $P_2^2$, and are therefore identified to a single blue vertex in the contracted graph. Likewise, all red vertices in $g_{24}$ are identified to a single red vertex in the contracted graph. For the specific case of $\ket{0}^{\otimes n}$, the two entropy vectors in $g_{24}$ correspond to completely unentangled states, or states which share an EPR pair among two qubits. 

As a group-theoretic object, the vertices of a contracted graph represent the equivalence classes of a double coset space, as defined in Eq.\ \eqref{DoubleCosetDefinition}. For the group $(HC)_{1,2}$ acting on $\Hil$, the subgroup
\begin{equation}
    (HP^2)_{1,2} \equiv \langle H_1,\,H_2,\,P_1^2,\,P_2^2 \rangle
\end{equation}
can never modify the entropy vector of any state. Accordingly, the $2$ vertices of the contracted graph in Figure \ref{G24WithContractedGraph} indicate the $2$ distinct equivalence classes in the double coset space $(\overline{HP^2})_{1,2} \backslash (\overline{HC})_{1,2} / \mathcal{S}_{\overline{HC}}(\ket{0}^{\otimes n})$. A representative element for each double coset equivalence class, shown in the above contracted graph, is $H_1$ (for blue) and $H_1C_{1,2}$ (for red).

\begin{figure}[h]
        \centering
        \includegraphics[width=15cm]{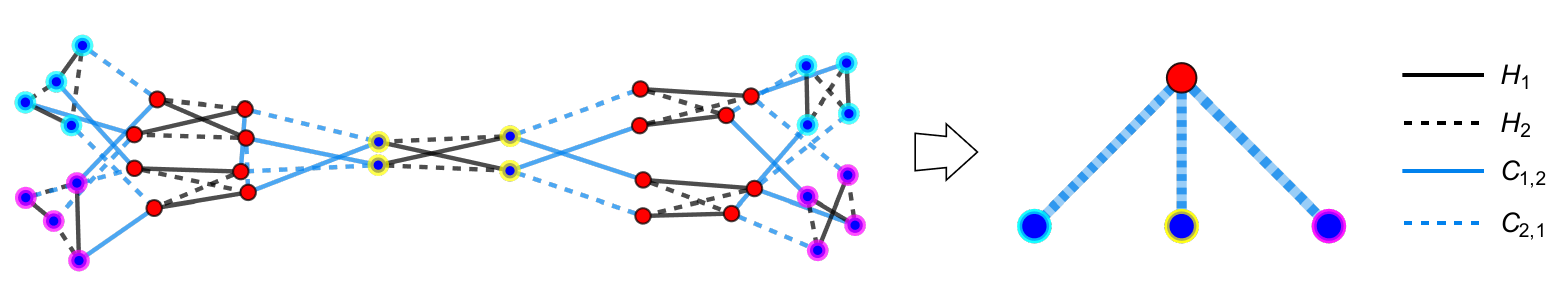}
        \caption{Reachability graph $g_{36}$ (left) and its contracted graph (right). All circuits are generated by the $4$ gates shown to the very right of the figure. The $g_{36}$ contracted graph contains $4$ vertices, but only ever realizes $2$ entropy vectors among those vertices. Different sets of blue vertices, highlighted in cyan, yellow, and magenta, identify respectively to the three blue vertices in the contracted graph. All red vertices in $g_{36}$ identify to a single red vertex in the contracted graph. Non-trivial entropy-preserving circuits, e.g.\ $(C_{i,j}H_j)^4$ from Eq.\ \eqref{HCLocalAction}, map vertices on opposite sides of $g_{36}$ to each other.}
        \label{G36WithContractedGraph}
    \end{figure}

Acting with the gates $H_1$ followed by $P_1$ on the state $\ket{0}^{\otimes n}$, that is
\begin{equation}
    \ket{\phi} = P_1H_1\ket{0}^{\otimes n},
\end{equation}
yields a state $\ket{\phi}$ with stabilizer group $\mathcal{S}_{\overline{HC}}(\ket{\phi})$, consisting of $32$ elements, which is not isomorphic to $\mathcal{S}_{\overline{HC}}(\ket{0}^{\otimes n})$. Consequently the state $\ket{\phi}$, as well as any other state with stabilizer group isomorphic to $\mathcal{S}_{\overline{HC}}(\ket{\phi})$, is not found on any $g_{24}$ graph. Instead, each state stabilized by $\mathcal{S}_{\overline{HC}}(\ket{\phi})$ resides on a reachability graph of $36$ vertices, which we term $g_{36}$, shown on the left of Figure \ref{G36WithContractedGraph}. In general, any state which is the product of a $2$-qubit stabilizer state and a generic $(n-2)$-qubit state will either have reachability graph $g_{24}$ or $g_{36}$.

The contracted graph of $g_{36}$, shown in the right panel of Figure \ref{G36WithContractedGraph}, contains $4$ vertices. All red vertices in $g_{36}$ identify to the same red vertex in the contracted graph. There are three distinct sets of blue vertices in $g_{36}$, highlighted with colors cyan, yellow, and magenta in Figure \ref{G36WithContractedGraph}, which identify to the three blue vertices in the contracted graph. All vertices highlighted by the same color in $g_{36}$ are connected by circuits which preserve the entropy vector.

The vertices of the $g_{36}$ contracted graph in Figure \ref{G36WithContractedGraph} represent the $4$ unique equivalence classes of the double coset space $(\overline{HP^2})_{1,2} \backslash \barHC / \mathcal{S}_{\overline{HC}}(\ket{\phi})$. A representative element for each equivalence class is $H_1$ (for cyan), $H_1C_{1,2}$ (for red), $H_1C_{1,2}C_{2,1}$ (for pink), and $H_1C_{1,2}H_2C_{1,2}$ (for yellow). Examining the vertex identifications in Figure  \ref{G36WithContractedGraph}, we again observe that the contraction map is not a quotient map on the original group. Vertex sets of different cardinalities in $g_{36}$ are identified together under this graph contraction, which cannot occur in a formal group quotient.

While the $g_{36}$ contracted graph contains four vertices, these vertices only ever realize two different entropy vector possibilities. Specifically, the two entropy vectors found on any $g_{36}$ graph are exactly the same as those found on the $g_{24}$ graph in Figure \ref{G24WithContractedGraph}. As we will show below, graph $g_{24}$ attaches to $g_{36}$ when we add phase gates back to our generating set. This connection of the $g_{24}$ and $g_{36}$ reachability graphs by local operations constrains the number of distinct entropy vectors that can be found on either graph.

\subsubsection{$\mathcal{C}_2$ Contracted Graphs of $g_{24}$ and $g_{36}$}

\begin{figure}[h]
        \centering
        \includegraphics[width=13cm]{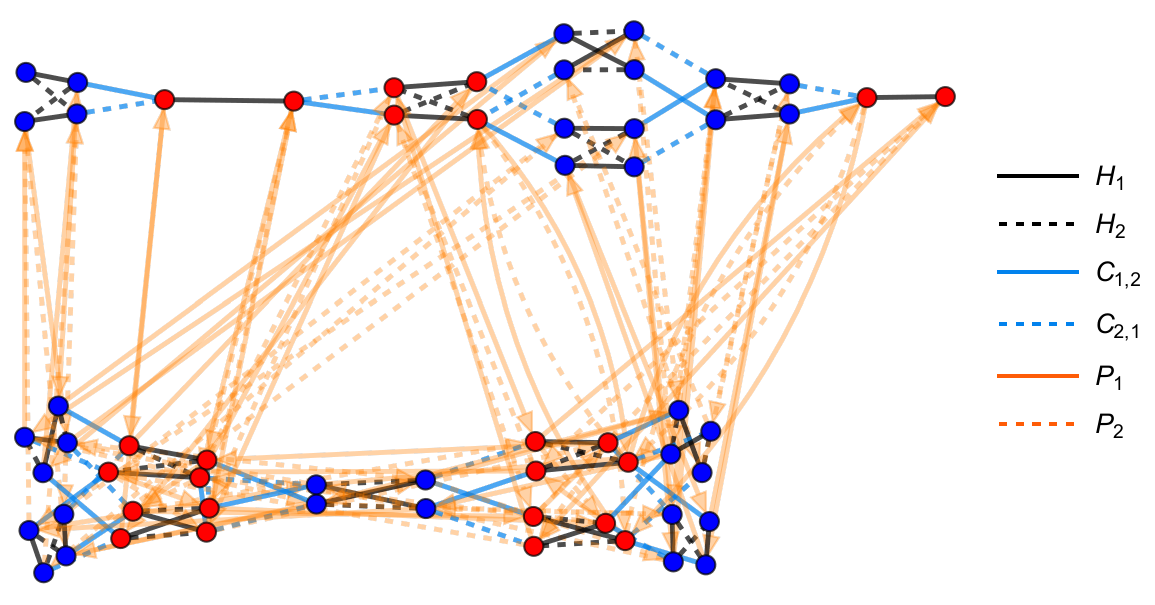}
        \caption{$\mathcal{C}_2$ Reachability graph for all states with $\mathcal{S}_{\overline{HC}}(\ket{0}^{\otimes n})$. All circuits are generated by the $6$ gates shown to the very right of the figure. This $60$-vertex reachability graph is the attachment of $g_{24}$ (Figure \ref{G24WithContractedGraph}) to $g_{36}$ (Figure \ref{G36WithContractedGraph}) by $P_1$ and $P_2$ gates. This reachability graph is likewise shared by all stabilizer product states.}
        \label{PhaseConnectedG24:G36}
    \end{figure}

We now analyze the full action of $\overline{\mathcal{C}}_2$ on states in a $g_{24}$ or $g_{36}$ reachability graph under $\barHC$. Acting with $\overline{\mathcal{C}}_2$ on any such state generates a reachability graph of $60$ vertices, which can be seen in Figure \ref{PhaseConnectedG24:G36}. This $60$-vertex reachability graph consists of a single copy of $g_{24}$ (top), attached to a single copy of $g_{36}$ (bottom) by sets of $P_1$ and $P_2$ edges.

Following the $P_1$ and $P_2$ edges in Figure \ref{PhaseConnectedG24:G36}, we can observe how vertices of a certain color connect to other vertices of the same color. Blue vertices in $g_{24}$ always connect to blue vertices in $g_{36}$, as is true for red vertices. Red vertices in $g_{36}$ may connect to other red vertices in $g_{36}$, or to red vertices in $g_{24}$. The three distinct batches of blue vertices in $g_{36}$, highlighted in Figure \ref{G36WithContractedGraph}, connect to each other via sequences of $H_1,\,H_2,\,P_1,$ and $P_2$, all of which leave the entropy vector unchanged. We can also directly observe circuits such as $\left(C_{1,2}H_2 \right)^4$, as in Eq.\ \eqref{HCLocalAction}, and verify that this sequence is indeed equivalent to the entropy-preserving $P_1^2$ operation.

\begin{figure}[h]
        \centering
        \includegraphics[width=13cm]{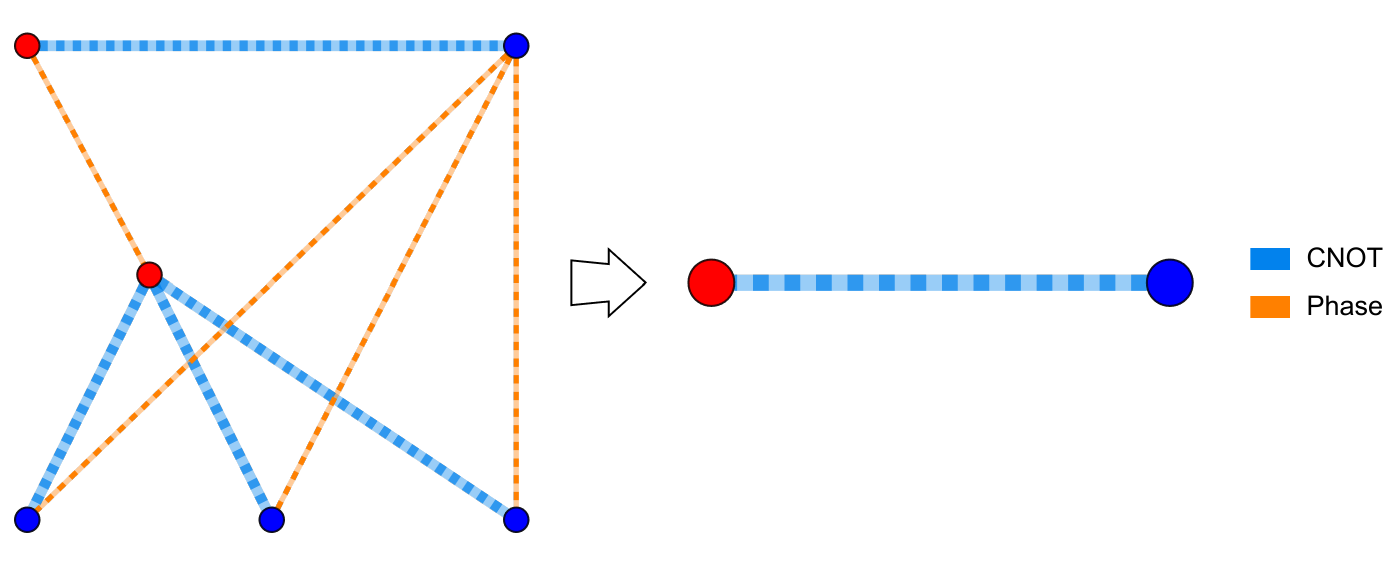}
        \caption{Contracted graph (right) of $\overline{\mathcal{C}}_2$ reachability graph in Figure \ref{PhaseConnectedG24:G36}. The left panel shows the contracted graphs of $g_{24}$ (top) and $g_{36}$ (bottom), connected by $P_1$ and $P_2$ circuits. Identifying vertices connected by phase edges quotients the left graph to the $2$-vertex contracted graph on the right. The $2$ vertices of this contracted graph represent the $2$ unique entropy vectors that can be found in the reachability graph in Figure \ref{PhaseConnectedG24:G36}.}
        \label{PhaseConnectedG24:g36ContractedGraph}
    \end{figure}

As before, we contract the $\mathcal{C}_2$ reachability graph in Figure \ref{PhaseConnectedG24:G36} by identifying vertices connected by entropy-preserving circuits. When performing this contraction on the full $\mathcal{C}_2$ graph we do not rely on any special operator relations, e.g.\ Eq.\ \eqref{HCLocalAction}, since we are identifying vertices connected by all $2$-qubit local operations, i.e.\ all operations built of $H_1,\,H_2,\,P_1,$ and $P_2$. The contracted graph of the $\mathcal{C}_2$ reachability graph in Figure \ref{PhaseConnectedG24:G36} is shown in the right panel of Figure \ref{PhaseConnectedG24:g36ContractedGraph}. The $2$ vertices in this contracted graph represent the $2$ equivalence classes in $(\overline{HP^2})_{1,2} \backslash \overline{\mathcal{C}}_2 / \mathcal{S}_{\overline{\mathcal{C}}_2}(\ket{0}^{\otimes n})$.

Figure \ref{PhaseConnectedG24:G36} depicts how sets of phase gates connect reachability graphs $g_{24}$ and $g_{36}$. Similarly, the left panel of Figure \ref{PhaseConnectedG24:g36ContractedGraph} shows how the respective contracted graphs of $g_{24}$ and $g_{36}$ are connected by sets of phase edges. The right panel of Figure \ref{PhaseConnectedG24:g36ContractedGraph} gives the final contracted graph after quotienting the $\mathcal{C}_2$ reachability graph in Figure \ref{PhaseConnectedG24:G36} by all entropy-preserving edges. The contracted graph has $2$ vertices, corresponding to the $2$ possible entropy vectors that can be found on any $\mathcal{C}_2$ reachability graph of the form shown in Figure \ref{PhaseConnectedG24:G36}. Furthermore, the $2$ vertices in the contracted explain why both graphs $g_{24}$ and $g_{36}$ individually only ever realize $2$ entropy vector colors among their vertices.

We examined the action of $\barHC$ and $\overline{\mathcal{C}}_2$ on $n$-qubit states with stabilizer group isomorphic to $\mathcal{S}_{\overline{HC}}(\ket{0}^{\otimes n})$ and $\mathcal{S}_{\overline{HC}}(P_1H_1\ket{0}^{\otimes n})$. We generated the reachability graphs for all states with both stabilizer groups, and quotiented each reachability graph by entropy-preserving operations to build the associated contracted graphs. The number of vertices in each contracted graph gave an upper bound on the number of different entropy vectors found in each reachability graph. Similarly, the edges in each contracted graph indicated the ways an entropy vector can change under all circuits comprising the reachability graph. We will now consider the reachability graphs of $n > 2$ qubit stabilizer states, where more-complicated entanglement structures can arise.

\subsection{Contracted Graphs of $g_{144}$ and $g_{288}$}

When we consider the action of $\barHC$ and $\overline{\mathcal{C}}_2$ on systems of $n > 2$ qubits, new reachability graph structures appear \cite{Keeler2022}. Additionally at $n > 2$ qubits, we observe new entanglement possibilities as well as new entropy vector colorings for reachability graphs. In this subsection, we define two new sets of stabilizer states which arise at $n=3$ qubits, defined by their stabilizer subgroup under $\barHC$ action. We build all reachability graphs and contracted graphs for these two families of states, and determine the bounds on entropy vector evolution in their respective reachability graphs. We then consider the full action of $\overline{\mathcal{C}}_2$ on these classes of states, and again build all reachability and contracted graphs.

\begin{figure}[h]
        \centering
        \includegraphics[width=14.5cm]{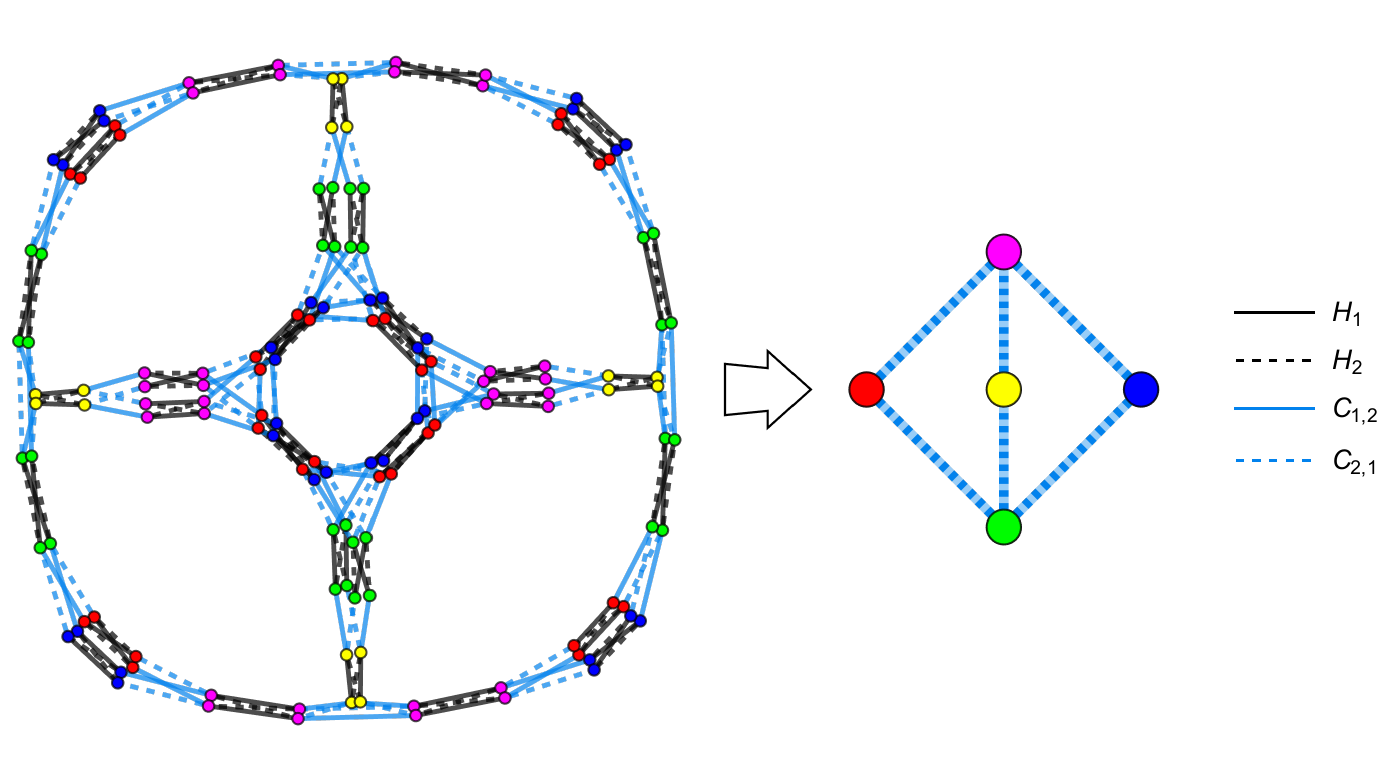}
        \caption{Reachability graph $g_{144}$ (left), and its associated contracted graph (right). All circuits are generated by the $4$ gates shown to the very right of the figure. The contracted graph contains $5$ vertices, corresponding to the $5$ unique entropy vectors that can be found on $g_{144}$. We depict a $g_{144}$ graph for the $6$-qubit state defined in Eq.\ \eqref{SixQubit144State}, as it contains the maximal number of $5$ entropy vectors among its vertices. Again we observe that certain circuits, e.g.\ Eq.\ \eqref{HCLocalAction}, do not modify entanglement and map vertices of the same color together. The specific entropy vectors shown are given in Table \ref{tab:g144g288EntropyVectorTable}.}
        \label{G144WithContractedGraph}
    \end{figure}

At three qubits, acting with $\barHC$ on certain stabilizer states produces an additional two reachability graphs beyond $g_{24}$ and $g_{36}$ discussed in the previous subsection. One new graph which arises at three qubits contains $144$ vertices, shown on the left of Figure \ref{G144WithContractedGraph}, and corresponds to states which are stabilized by $8$ elements in $\barHC$. One example of a state with $g_{144}$ reachability graph is the $3$-qubit GHZ state $\ket{GHZ}_3 \equiv \ket{000} + \ket{111}$. The graph $g_{144}$ is shared by all states with a stabilizer subgroup isomorphic to $\mathcal{S}_{\overline{HC}}(\ket{GHZ}_3)$. For reasons we will explain in a moment, Figure \ref{G144WithContractedGraph} depicts the specific reachability graph for the $6$-qubit state defined in Eq.\ \eqref{SixQubit144State}.

The contracted graph of $g_{144}$, shown on the right of Figure \ref{G144WithContractedGraph}, contains $5$ vertices. These $5$ vertices represent the $5$ unique entropy vectors that can be found on any $g_{144}$ reachability graph. While the graph $g_{144}$ is first observed among $3$-qubit systems, we do not find a maximal coloring of $g_{144}$, i.e.\ a copy of $g_{144}$ with $5$ different entropy vectors, until $6$ qubits. The specific graph shown in Figure \ref{G144WithContractedGraph} corresponds to the orbit of the $6$-qubit state defined in Eq.\ \eqref{SixQubit144State}, which we choose precisely because its $g_{144}$ graph displays the maximum allowable entropic diversity. The specific entropy vectors corresponding to the colors seen in Figure \ref{G144WithContractedGraph} can be found in Table \ref{tab:g144g288EntropyVectorTable} of Appendix \ref{EntropyVectorTables}.

\begin{figure}[h]
        \centering
        \includegraphics[width=15.1cm]{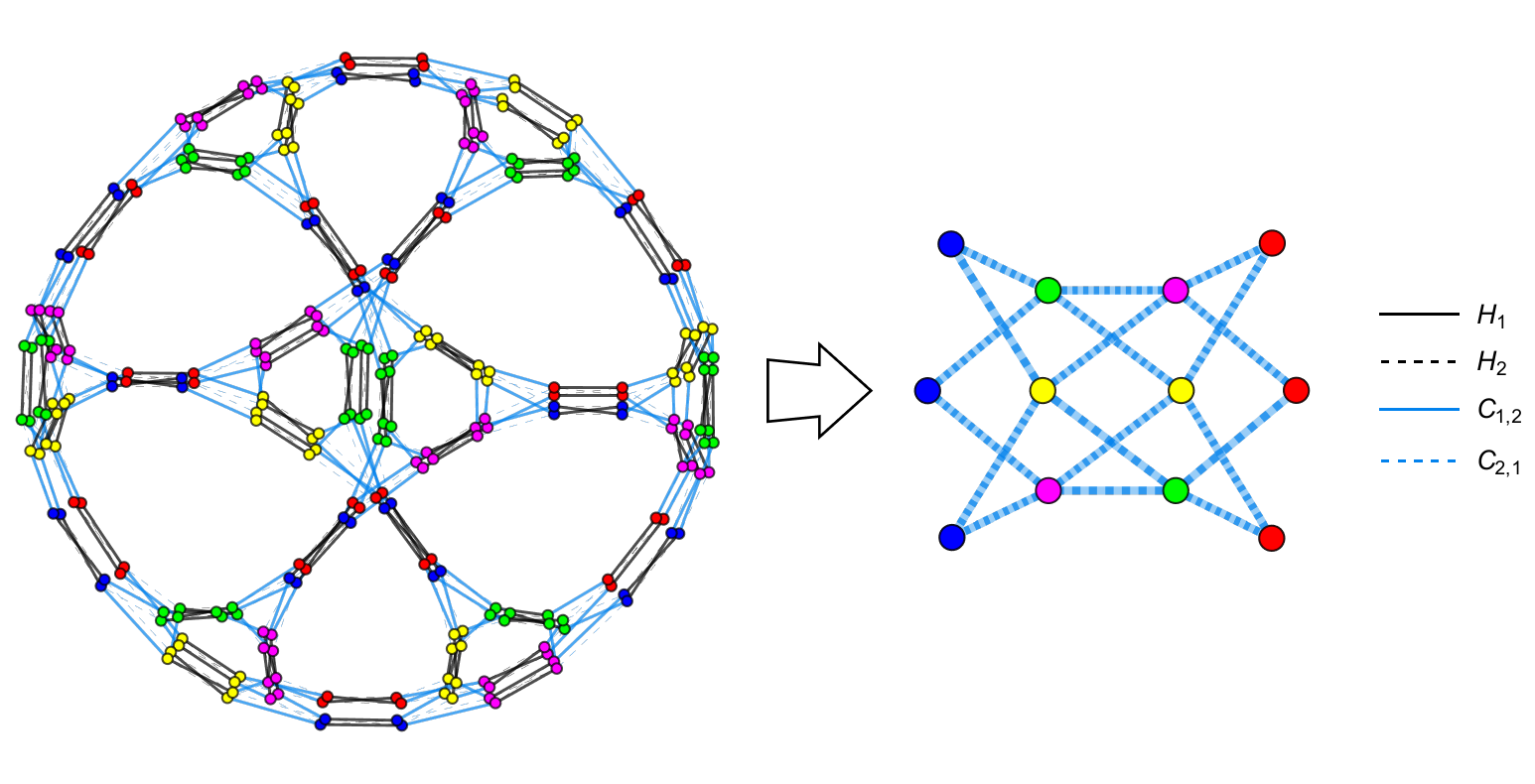}
        \caption{Reachability graph $g_{288}$, and its contracted graph, for $6$-qubit state stabilized by Eq.\ \eqref{g288StabGroup}. All circuits are generated by the $4$ gates shown to the very right of the figure. While the $g_{288}$ contracted graph has $12$ vertices, we only ever witness $5$ entropy vectors among those vertices. The specific entropy vectors depicted are the same as those in Figure \ref{G144WithContractedGraph}, and can be found in Table \ref{tab:g144g288EntropyVectorTable}.}
        \label{G288WithContractedGraph}
    \end{figure}

Also beginning at three qubits, we witness a stabilizer state reachability graph with $288$ vertices, which we denote $g_{288}$. States with reachability graph $g_{288}$ are stabilized by $4$ elements of $\barHC$, specifically by a subgroup whose equivalence classes can be represented by
\begin{equation}\label{g288StabGroup}
    \{\mathbb{1},\, H_2(C_{1,2}H_1)^4,\, (C_{1,2}H_1)^4H_2,\, \left((C_{1,2}H_1)^3C_{1,2}H_2\right)^2\}.
\end{equation}
The left panel of Figure \ref{G288WithContractedGraph} depicts a $g_{288}$ reachability graph, specifically for a $6$-qubit state stabilized by the group in Eq.\ \eqref{g288StabGroup}.

The $g_{288}$ contracted graph shown in the right panel of Figure \ref{G288WithContractedGraph} contains $12$ vertices, which provides a weak upper bound on the number of entropy vectors that can be found on any $g_{288}$ graph. However, for reasons we will soon explain, the $12$ vertices of this contracted graph are only ever colored by $5$ different entropy vectors. The specific $5$ entropy vectors shown in Figure \ref{G288WithContractedGraph} are exactly those seen in Figure \ref{G144WithContractedGraph}, and are defined in Table \ref{tab:g144g288EntropyVectorTable}. Similar to the case of $g_{144}$ in Figure \ref{G144WithContractedGraph}, the graph $g_{288}$ is first observed among $3$-qubit systems, but only witnesses a maximal coloring beginning at $n \geq 6$ qubits. 

\begin{figure}[h]
        \centering
        \includegraphics[width=14cm]{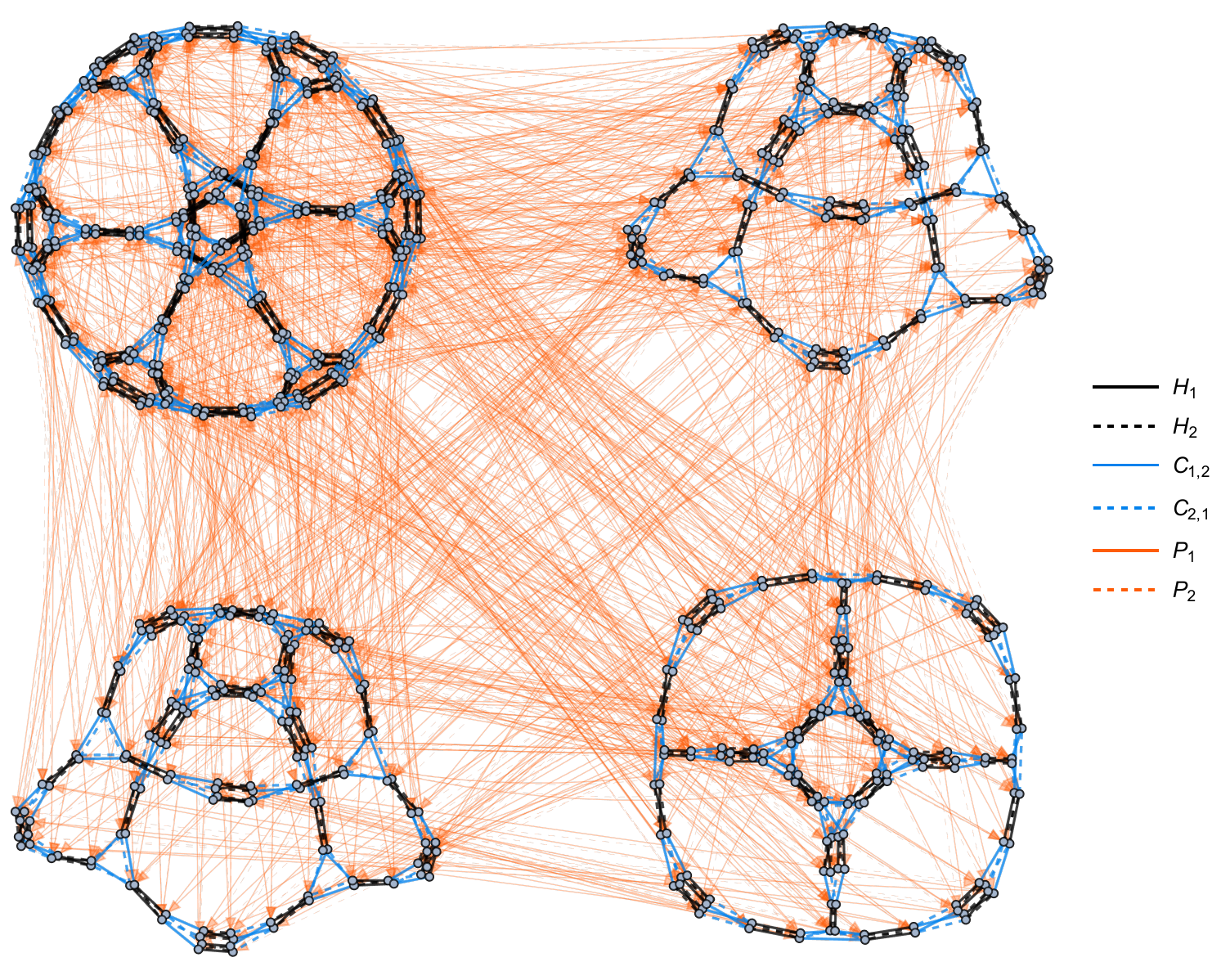}
        \caption{Reachability graph for states in $g_{144}$ and $g_{288}$ graphs, under the full action of $\overline{\mathcal{C}}_2$. All circuits are generated by the $6$ gates shown to the very right of the figure. This $768$-vertex graph is composed of $3$ copies of $g_{144}$ and a single $g_{288}$. Each copy of $g_{144}$ is isomorphic, despite the layouts being slightly different above. The graph connectivity constrains the diversity of entropy vectors which can be found on any single $g_{144}$ and $g_{288}$ graph. For clarity we choose not to color vertices by their entropy vector here.}
        \label{HhCcPhaseOverlay}
    \end{figure}

We now consider the full action of $\overline{\mathcal{C}}_2$ on states with a $g_{144}$ or $g_{288}$ reachability graph, returning $P_1$ and $P_2$ to our generating set. Every state in a $g_{144}$ and $g_{288}$ reachability graph under $\barHC$ is stabilized by $15$ elements of the full group $\overline{\mathcal{C}}_2$. The orbit of all such states under $\overline{\mathcal{C}}_2$ therefore contains $768$ states, and the associated $768$-vertex reachability graph is shown in Figure \ref{HhCcPhaseOverlay}. The orange edges in the reachability graph, which correspond to $P_1$ and $P_2$ gates, illustrate specifically how three different copies of $g_{144}$ attach to a single copy of $g_{288}$ under phase operations.

\begin{figure}[h]
        \centering
        \includegraphics[width=14cm]{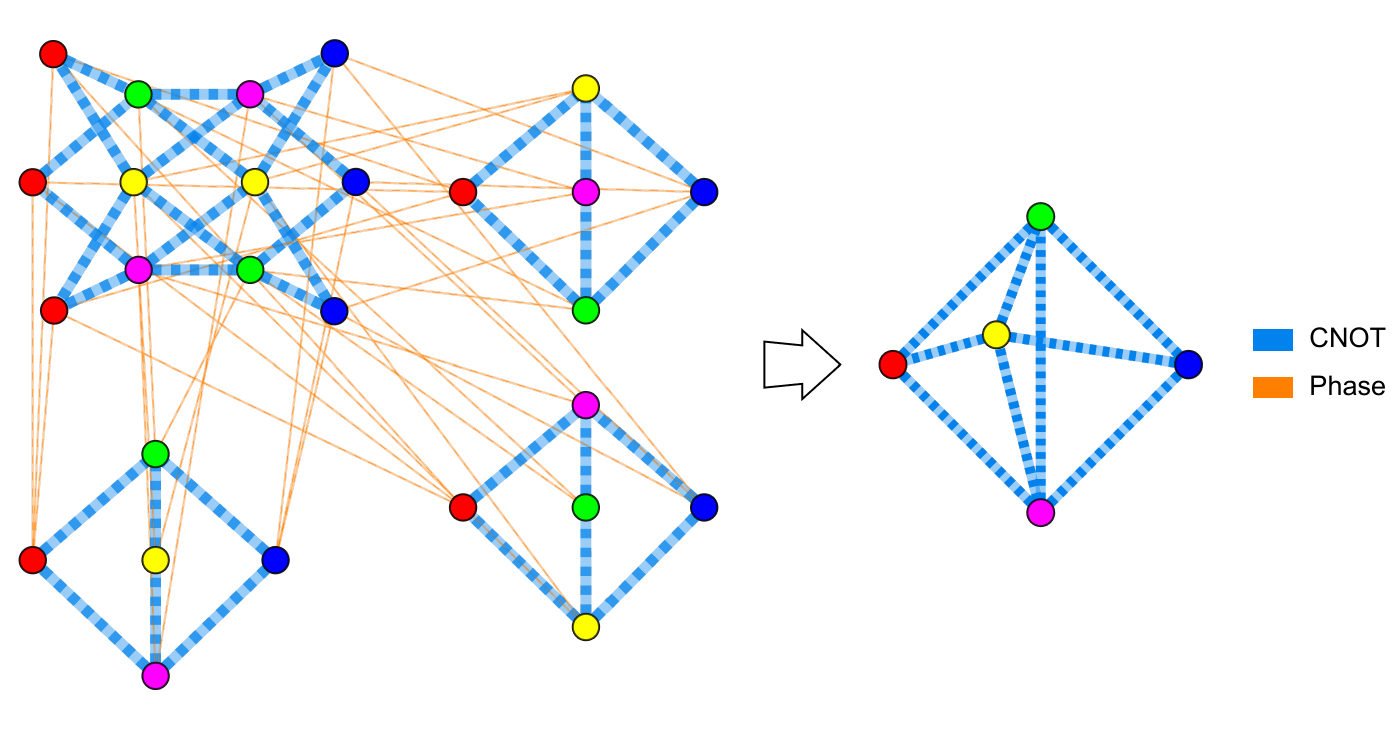}
        \caption{Contracted graph of $\mathcal{C}_2$ reachability graph from Figure \ref{HhCcPhaseOverlay}. The left panel depicts the individual contracted graphs of the $3$ $g_{144}$ graphs attached to a single $g_{288}$ graph. The right panel shows the final contracted graph, with $5$ vertices, and explains why we only ever find $g_{288}$ and $g_{144}$ graphs with $5$ different entropy vectors (given in Table \ref{tab:g144g288EntropyVectorTable}).}
        \label{PhaseConnectedg144_288ContractedGraph}
    \end{figure}

The contracted graphs for each $g_{144}$ and $g_{288}$ in Figure \ref{HhCcPhaseOverlay} are compiled in the left panel of Figure \ref{PhaseConnectedg144_288ContractedGraph}. Each of the three copies of $g_{144}$ contracts to a $5$-vertex graph that is isomorphic to Figure \ref{G144WithContractedGraph}, while the single copy of $g_{288}$ contracts to the $12$-vertex graph seen in Figure \ref{G288WithContractedGraph}. These four contracted graphs attach to each other under phase operations, adding connections which do not change a state's entropy vector. The final contracted graph of Figure \ref{HhCcPhaseOverlay} is shown on the right of Figure \ref{PhaseConnectedg144_288ContractedGraph}, and only has $5$ vertices.

The full $\mathcal{C}_2$ contracted graph in Figure \ref{PhaseConnectedg144_288ContractedGraph} is almost identical to the $g_{144}$ contracted graph in Figure \ref{G144WithContractedGraph}, but with an additional edge connecting two of the vertices. Since every $g_{288}$ attaches to $3$ copies of $g_{144}$ by phase gates, which do not modify entanglement, the maximum number of entropy vectors on any $g_{288}$ is bounded by the entropic coloring of each $g_{144}$ it connects to. This connectivity explains why we only observe at most $5$ entropy vectors on any $g_{288}$ graph, as can be seen in Figure \ref{G288WithContractedGraph}.

Figure \ref{PhaseConnectedg144_288ContractedGraph} depicts a symmetry between red and blue vertices which corresponds to an equivalence of these two entropy vectors under an exchange of the first two qubits. We likewise observe a symmetry between green, yellow, and magenta vertices, reflecting the three ways to divide the $4$-qubit subsystem $CDEO$ into two groups of two qubits each. For each $g_{144}$ contracted graph in Figure \ref{PhaseConnectedg144_288ContractedGraph}, the middle vertex corresponds to the entropy vector that occurs the fewest number of times, specifically $16$ times, in each respective $g_{144}$ reachability graph. We again observe that the contraction procedure generates a double coset space, rather than a group quotient, since the resulting equivalence classes have different cardinalities.

In this subsection we built contracted graphs for the stabilizer state reachability graphs $g_{144}$ and $g_{288}$, corresponding to states which are stabilized by $4$ and $8$ elements of $\barHC$ respectively. We showed how the contracted graph for $g_{144}$, with $5$ vertices, and the contracted graph for $g_{288}$, with $12$ vertices, both witness a maximum of $5$ different entropy vectors. This constraint on the number of different entropy vectors, perhaps surprising in the case of $g_{288}$, can be understood by considering the full action of $\overline{\mathcal{C}}_2$, which attaches three copies of $g_{144}$ to $g_{288}$ by phase operations. The number of entropy vectors found on any $g_{288}$ reachability graph is bounded by the number of entropy vectors found on each of the $g_{144}$ graphs to which it attaches, since $P_1$ and $P_2$ cannot modify entanglement. In the next subsection we consider the action of $\barHC$ and $\overline{\mathcal{C}}_2$ on generic quantum states, which allows us to extend our analysis beyond the stabilizer states. 

\subsection{Contracted Graphs of $g_{1152}$ and Full $\mathcal{C}_2$}

\begin{figure}[h]
        \centering
        \includegraphics[width=15cm]{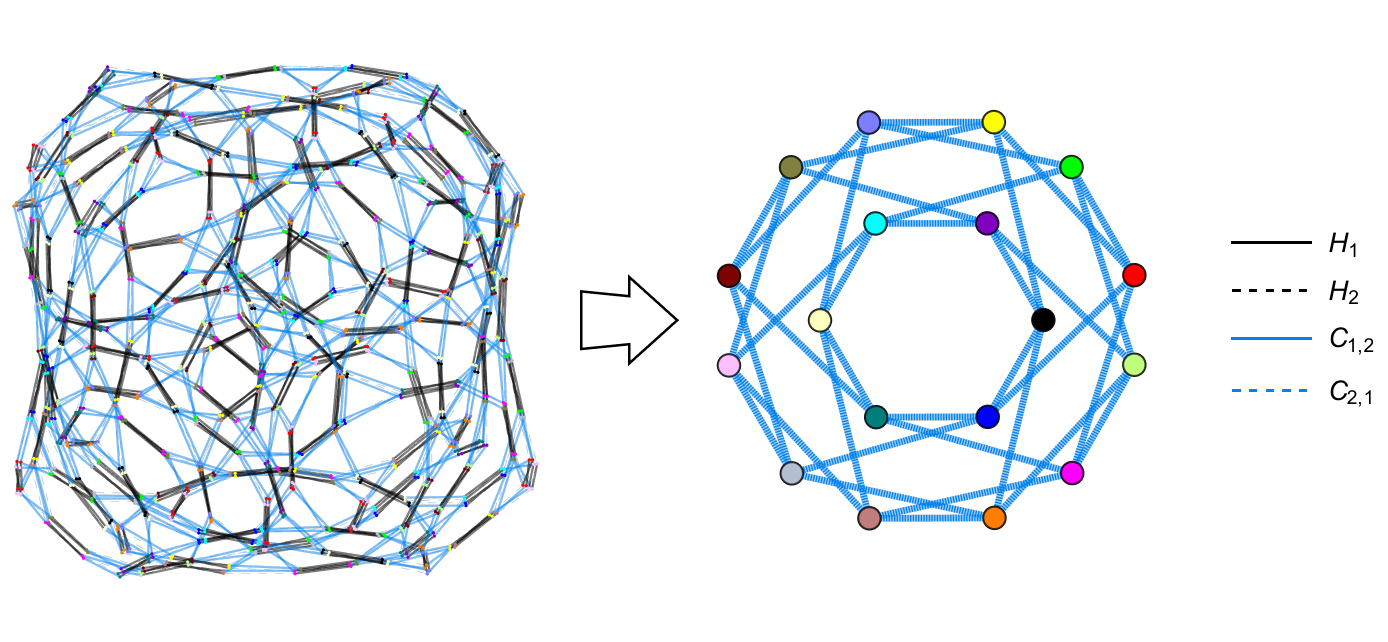}
        \caption{Reachability graph $g_{1152}$ (left) and its contracted graph (right). The graph $g_{1152}$ is shared by all stabilizer states stabilized by only $\mathbb{1} \in \barHC$, as well as generic quantum states. In this Figure, we illustrate an example $g_{1152}$ for the $8$-qubit state in Eq. \eqref{EightQubitState}, where the contracted graph achieves a maximal coloring of $18$ different entropy vectors (given in Figure \ref{EightQubitEntropyVectors}).}
        \label{G1152WithContractedGraph}
    \end{figure}

We now study the generic $\HC$ reachability graph for any quantum state stabilized by only the identity in $\HC$. In \cite{Keeler2022} Sections 4--5, we demonstrate that there exist only $5$ unique cosets $(HC)_{1,2}/\textnormal{Stab}_{\psi}$ (up to isomorphism), where $\psi$ is a stabilizer state. These cosets are represented by the reachability graphs in Figures \ref{G24WithContractedGraph}, \ref{G36WithContractedGraph}, \ref{G144WithContractedGraph}, \ref{G288WithContractedGraph}, and \ref{G1152WithContractedGraph}. At the group level, the number of vertices in each graph is fixed to be an integer divisor of the overall $1152$ element group $(HC)_{1,2}$. The specific factors of $1152$ which correspond to the order of stabilizer groups of stabilizer states are $1,\, 4,\, 8,\, 32,$ and $48$, producing the specific reachability graphs above. A reachability graph of $1152$ vertices corresponds to a state stabilized by only $1$ element, the identity, which constitutes the largest possible reachability graph under $(HC)_{1,2}$. For stabilizer state systems, this final $\HC$ reachability graph structure arises at $n \geq 4$ qubits. The reachability graph, which we term $g_{1152}$, contains $1152$ vertices and is shown on the left of Figure \ref{G1152WithContractedGraph}.

The contracted graph of $g_{1152}$, shown in the right panel of Figure \ref{G1152WithContractedGraph}, contains $18$ vertices. These $18$ vertices indicate the maximum number of unique entropy vectors that can be generated for any quantum state using only operations in $\HC$. The $g_{1152}$ contracted graph is symmetric, and achieves a maximal coloring at $8$ qubits. The specific instance of $g_{1152}$ in Figure \ref{G1152WithContractedGraph} corresponds to the $8$-qubit state given in Eq. \eqref{EightQubitState}, for which the entropy vectors are given in Table \ref{EightQubitEntropyVectors}.

\begin{figure}[h]
        \centering
        \includegraphics[width=15cm]{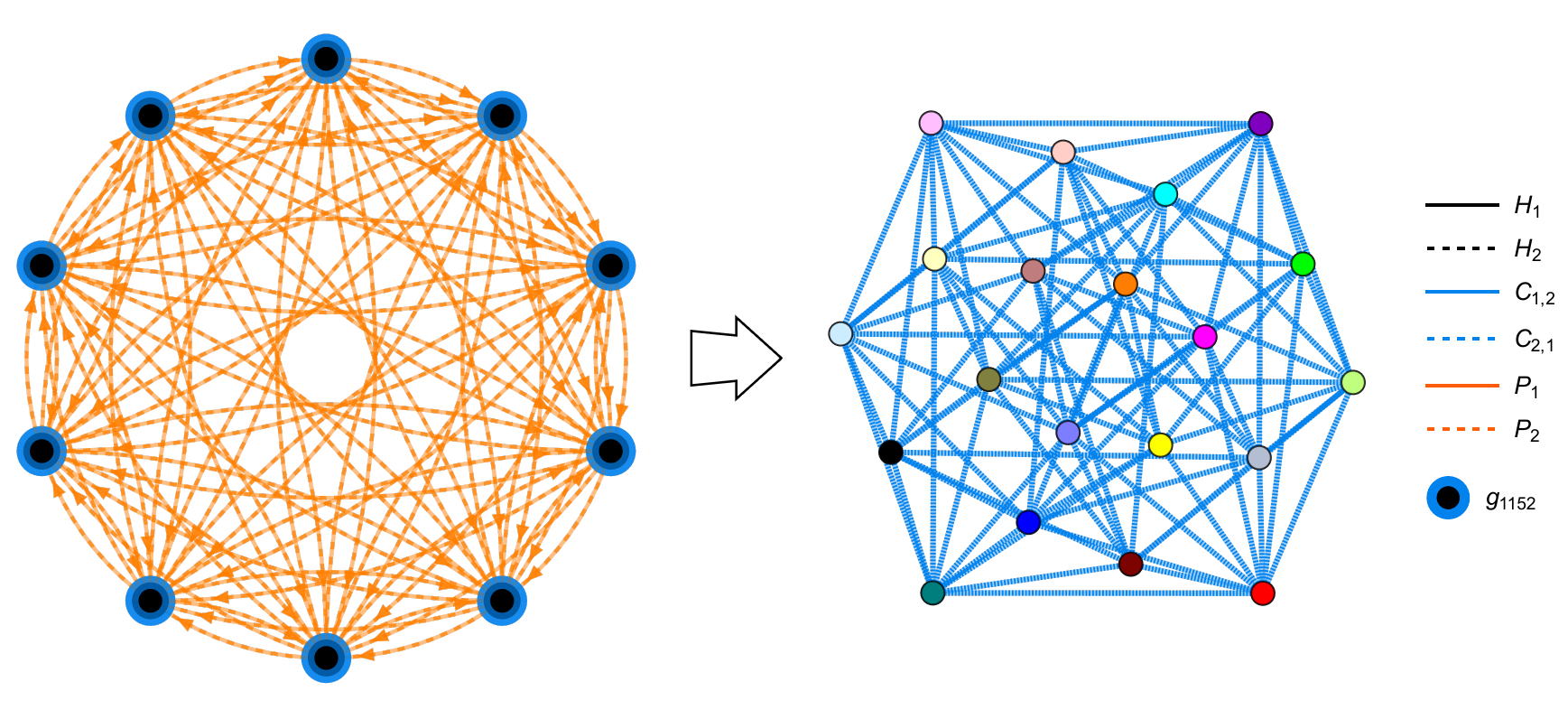}
        \caption{The full $\mathcal{C}_2$ reachability graph (left) with $11520$ vertices. We present this reachability graph as a collection of attached $g_{1152}$ graphs, illustrating how $\HC$ reachability graphs connect via $P_1$ and $P_2$ gates. We also remove all loops in the $\mathcal{C}_2$ reachability graph, i.e.\ all phase edges which map a copy of $g_{1152}$ to itself. The contracted graph of the $\mathcal{C}_2$ reachability graph is given to the right, and has $20$ vertices. These $20$ vertices give an upper bound on the number of distinct entropy vectors that can be reached by applying any sequence of $2$-qubit operations on any quantum state.}
        \label{FullC2WithContractedGraph}
    \end{figure}

The phase-quotiented $2$-qubit Clifford group $\overline{\mathcal{C}}_2$ is composed of $11520$ elements. A generic quantum state will only be stabilized by $\mathbb{1} \in \overline{\mathcal{C}}_2$, and therefore has an orbit of $11520$ states under $\overline{\mathcal{C}}_2$ action. Every state in an $11520$-vertex reachability graph under $\overline{\mathcal{C}}_2$ will trivially lie in a $g_{1152}$ graph under $\barHC$, however, the converse%
\footnote{Since any state in an $11520$-vertex graph under $\overline{\mathcal{C}}_2$ is stabilized by only the identity, each state will likewise be stabilized by only the identity in $\barHC$. There exist states, however, which are stabilized by only the identity in $\barHC$, but can be transformed under phase gates into states stabilized by more than one element of $\barHC$. Subsection \ref{DickeStateSubsection} discusses two classes of Dicke states which demonstrate this counterexample. \label{C2Footnote}} %
is not always true. We display the full $\mathcal{C}_2$ reachability graph, in a compressed format, to the left of Figure \ref{FullC2WithContractedGraph}. Each vertex in the left panel of Figure \ref{FullC2WithContractedGraph} represents a distinct copy of $g_{1152}$ from Figure \ref{G1152WithContractedGraph}. Each of the $10$ copies of $g_{1152}$ attaches to every other $g_{1152}$ via $P_1$ and $P_2$ gates.

The contracted graph%
\footnote{Since all states in the $11520$-vertex reachability graph are stabilized by only $\mathbb{1} \in \overline{\mathcal{C}}_2$, and since $\langle \mathbb{1} \rangle$ is normal in $\overline{\mathcal{C}}_2$, the object $\overline{\mathcal{C}}_2/\langle \mathbb{1} \rangle$ defines a formal group quotient on $\overline{\mathcal{C}}_2$. Consequently, the contracted graph to the right of Figure \ref{FullC2WithContractedGraph} actually represents the right coset space $\overline{\mathcal{C}}_2 \backslash \overline{\langle H_1,\,H_2,\,P_1,\,P_2 \rangle}$, as opposed to a double coset space.} %
of the $11520$-vertex $\mathcal{C}_2$ reachability graph contains $20$ vertices, and is shown on the right of Figure \ref{FullC2WithContractedGraph}. This contracted graph is symmetric, i.e. it is both vertex and edge transitive, and the $20$ entropy vectors shown in Figure \ref{FullC2WithContractedGraph} are given in Table \ref{EightQubitEntropyVectors}. Since we are considering the full action of $\overline{\mathcal{C}}_2$, the $20$ vertices in this contracted graph constrain the number of entropy vectors that can be generated by any $2$-qubit Clifford circuit. Otherwise stated, given a generic quantum state with arbitrary entanglement structure, any unitary composed of $2$-qubit Clifford gates can maximally achieve $20$ distinct entropy vectors.

In the remainder of the section we extend our discussion beyond stabilizer states, examining contracted graphs for non-stabilizer Dicke states under $\barHC$ and $\overline{\mathcal{C}}_2$ action. We also derive a general upper bound for the number of entropy vectors that can be achieved under any $n$-qubit Clifford circuit, for arbitrary $n$.

\subsection{Non-Stabilizer State Contracted Graphs}\label{DickeStateSubsection}

Papers \cite{Keeler:2023xcx,Munizzi:2023ihc} showed that certain non-stabilizer states can have non-trivial stabilizer subgroups, i.e.\ they are stabilized by more than just the identity, under the action of $\overline{\mathcal{C}}_n$. One class of states in particular, the set of $n$-qubit Dicke states \cite{nepomechie2023qudit}, always admits a non-trivial $\overline{\mathcal{C}}_n$ stabilizer group. In this subsection, we discuss all $\HC$ and $\mathcal{C}_2$ reachability graphs for Dicke states and construct their associated contracted graphs. We use the contracted graphs to bound the number of possible entropy vectors that can be generated in Dicke state systems under Clifford group action \cite{Schnitzer:2022exe,Munizzi:2023ihc}.

Each $n$-qubit Dicke state $\ket{D^n_k}$ is defined as an equal-weight superposition over all $n$-qubit states of a fixed Hamming%
\footnote{In the bit-string representation of a quantum register, Hamming weight simply counts the number of $1$s.} %
weight. Using the $n$-qubit states $\{\ket{b}\}$, where $b$ denotes some binary string of length $2^n$, we construct $\ket{D^n_k}$ as the state
\begin{equation}
    \ket{D^n_k} \equiv \binom{n}{k}^{-1/2} \sum_{b \in \{0,1\}^n,\,h(b)=k} \ket{b},
\end{equation}
where $h(b) = k$ denotes the fixed Hamming weight of $b$. Some examples of Dicke states include
\begin{equation}
\begin{split}
        \ket{D^2_1} &= \frac{1}{\sqrt{2}}\left(\ket{01} + \ket{10}\right),\\
        \ket{D^4_2} &= \frac{1}{\sqrt{6}}\left(\ket{1100} + \ket{1010} + \ket{1001} + \ket{0110} + \ket{0101} + \ket{0011}\right).
\end{split}
\end{equation}
Dicke states of the form $\ket{D^n_1}$ are exactly the non-biseparable $n$-qubit $W$-states, while $\ket{D^n_n}$ are the computational basis states $\ket{1}^{\otimes n}$.

For $n \geq 3$ qubits, the state $\ket{D^n_1}$ is not a stabilizer state. Regardless, each $\ket{D^n_k}$ is stabilized by a subset of $\overline{\mathcal{C}}_n$ that contains more than just the identity. When considering the action of $\overline{\mathcal{C}}_2$ on $\ket{D^n_k}$, states of the form $\ket{D^n_1}$ and $\ket{D^n_{n-1}}$ share one particular set of stabilizers, while those of the form $\ket{D^n_k}$ with $1 < k < n-1$ share another. We discuss both cases below. 

Dicke states of the form $\ket{D^n_1}$ and $\ket{D^n_{n-1}}$ are not stabilizer states for all $n \geq 3$. However, both $\ket{D^n_1}$ and $\ket{D^n_{n-1}}$ are stabilized by a $4$-element subgroup%
\footnote{These elements are actually coset representatives of each equivalence class in the stabilizer group. There is a more compact representation of this stabilizer group using $CZ$ gates (see also \cite{Latour:2022gsf}), which can be written $\mathcal{S}_{\overline{HC}}(\ket{D^n_1}) = \{\mathbb{1},\,CZ_{1,2},\,C_{1,2}C_{2,1}C_{1,2},\,CZ_{1,2}C_{1,2}C_{2,1}C_{1,2}\}$.} %
of $\overline{\mathcal{C}}_2$, specifically
\begin{equation}\label{WStateStabGroup}
\begin{split}
    \mathcal{S}_{\overline{HC}}(\ket{D^n_1}) &= \{\mathbb{1},\,H_2C_{1,2}H_2,\,C_{1,2}C_{2,1}C_{1,2},\,H_2C_{1,2}H_2C_{1,2}C_{2,1}C_{1,2}\},\\
    &= \mathcal{S}_{\overline{HC}}(\ket{D^n_{n-1}}).
\end{split}
\end{equation}
Furthermore, we note that the subgroup in Eq.\ \eqref{WStateStabGroup} is contained in $\barHC$. Therefore the left coset space $\barHC / \mathcal{S}_{\overline{HC}}(\ket{D^n_1})$ contains $288$ elements.

\begin{figure}[h]
        \centering
        \includegraphics[width=15cm]{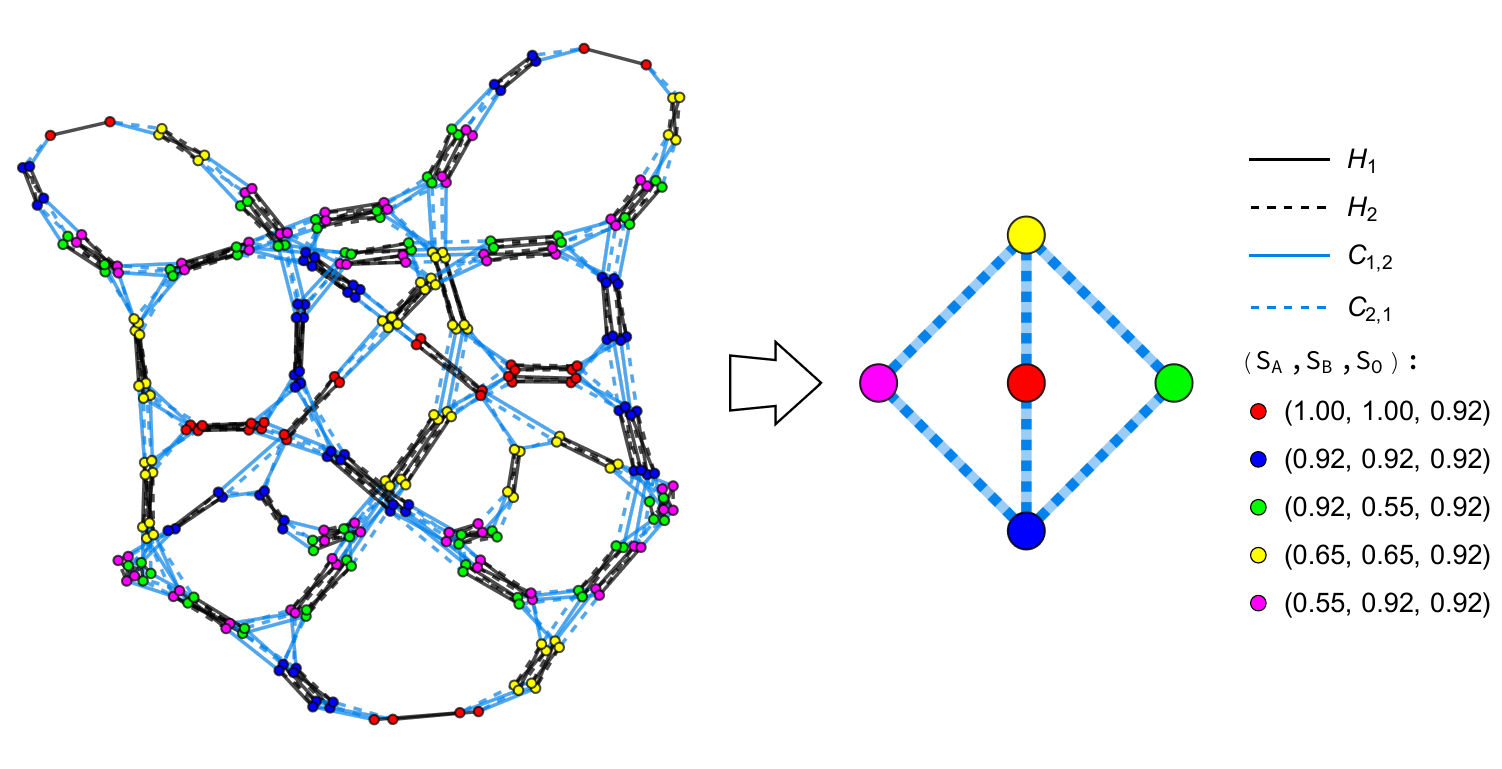}
        \caption{Reachability graph $g_{288^*}$ (left) for $\ket{D^3_1}$ under the action of $\barHC$. The graph $g_{288^*}$ has different topology than the $g_{288}$ graph for stabilizer states. The $g_{288^*}$ contracted graph (right) has $5$ vertices, and is isomorphic to the stabilizer state contracted graph of $g_{144}$ from Figure \ref{G144WithContractedGraph}. The exact, rather than numerical, values of the $5$ entropy vectors given in the legend are shown in Table \ref{tab:WStateEntropyVectorTable}.}
        \label{WStateG288WithContractedGraph}
    \end{figure}

The reachability graph for all $\ket{D^n_1}$ and $\ket{D^n_{n-1}}$, which we denote $g_{288^*}$, has $288$ vertices, as dictated by the order of $\mathcal{S}_{\overline{HC}}(\ket{D^n_1})$ in Eq.\ \eqref{WStateStabGroup}. While the graph $g_{288^*}$ has the same number of vertices as the $g_{288}$ graph for stabilizer states, shown in Figure \ref{G288WithContractedGraph}, its topology is distinct from $g_{288}$ and the two graphs are not isomorphic. Graphs with the topology of $g_{288^*}$ are never observed among stabilizer states, and provide an example of non-stabilizer states that are stabilized by more than just the identity in $\overline{\mathcal{C}}_2$. The left panel of Figure \ref{WStateG288WithContractedGraph} depicts an example of $g_{288^*}$, specifically for the state $\ket{D^3_1}$.

The contracted graph of $g_{288^*}$ has $5$ vertices, and is shown on the right of Figure \ref{WStateG288WithContractedGraph}. While the reachability graph $g_{288}$ for stabilizer states has a contracted graph of $12$ vertices, the distinct connectivity of $g_{288^*}$ yields a smaller contracted graph. Interestingly, the $g_{288^*}$ contracted graph is isomorphic to the $g_{144}$ contracted graph seen in Figure \ref{G144WithContractedGraph}. There are $5$ possible entropy vectors found on any $g_{288^*}$, and the graph achieves a maximal coloring beginning at $3$ qubits.

\begin{figure}[h]
        \centering
        \includegraphics[width=15cm]{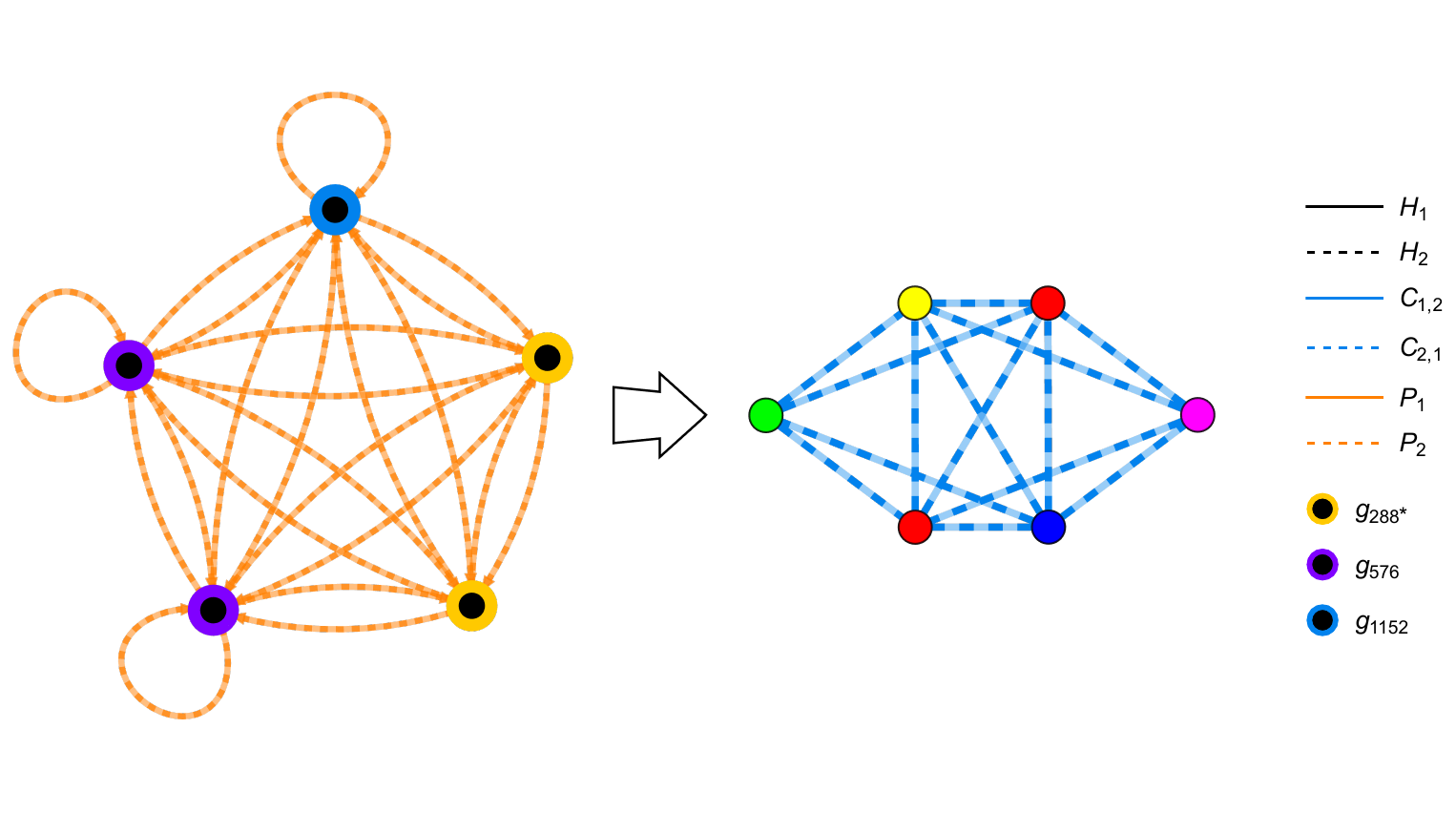}
        \caption{Reachability graph (left) of $\ket{D^3_1}$ under the full action of $\overline{\mathcal{C}}_2$, containing $2880$ vertices. We illustrate this reachability graph with vertices representing graphs $g_{288^*}, g_{576},$ and $g_{1152}$ to illustrate the connectivity of certain $\HC$ reachability graphs under phase gates. The right panel of the Figure depicts the associated contracted for the $\mathcal{C}_2$ reachability graph, which contains $6$ vertices.}
        \label{PhaseConnectedWG288ContractedGraph}
    \end{figure}

\begin{figure}[h]
        \centering
        \includegraphics[width=15cm]{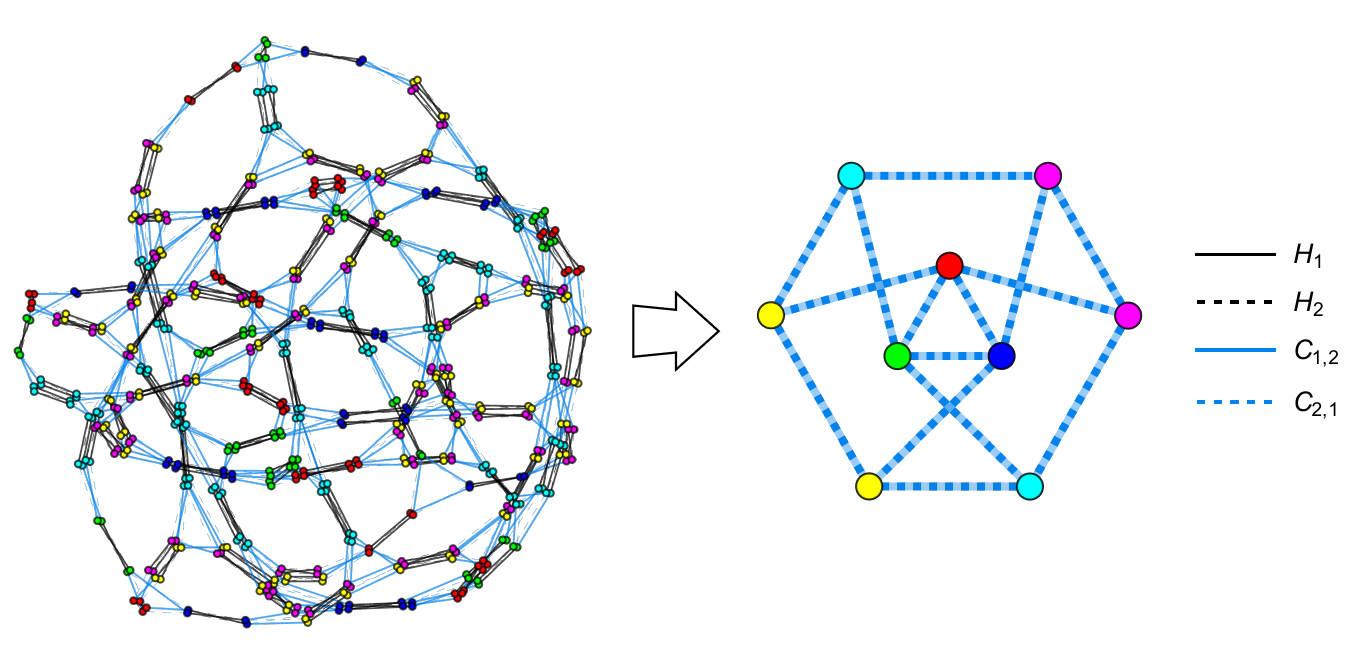}
        \caption{The $g_{576}$ reachability graph (left) for $\ket{D^4_2}$ under $\barHC$ action. Graphs of $576$ vertices are never observed among stabilizer states under $\barHC$ action. The graph $g_{576}$ contracts to a graph of $9$ vertices under entropy-preserving operations, with $6$ different entropy vectors among those vertices. The $6$ entropy vectors found in this contracted graph are given in Table \ref{tab:D42EntropyVectors}.}
        \label{G576WithContractedGraph}
    \end{figure}

The orbit of $\ket{D^n_1}$ and $\ket{D^n_{n-1}}$ under the full group $\overline{\mathcal{C}}_2$ reaches $2880$ states, generating a reachability graph of $2880$ vertices. The left panel of Figure \ref{PhaseConnectedWG288ContractedGraph} illustrates this $2880$-vertex reachability graph for the state $\ket{D^3_1}$, which is comprised of several attached copies of $\HC$ reachability graphs. For clarity, we allow each vertex of the $2880$-vertex reachability graph to represent graphs $g_{288^*},\,g_{576}$ (introduced later in Figure \ref{G576WithContractedGraph}), and $g_{1152}$, focusing on the connectivity between different $\barHC$ orbits under $P_1$ and $P_2$ operations.

The $\mathcal{C}_2$ reachability graph in Figure \ref{PhaseConnectedWG288ContractedGraph} is built of $2$ attached copies of $g_{288^*}$, $2$ copies of $g_{576}$, and a single $g_{1152}$. Every state in this $2880$-vertex reachability graph is stabilized by $4$ elements of $\overline{\mathcal{C}}_2$. Certain states, such as $\ket{D^n_1}$ and $\ket{D^n_{n-1}}$, are stabilized by a $4$-element subgroup of $\overline{\mathcal{C}}_2$ which is also completely contained within $\barHC$, as shown in Eq.\ \eqref{WStateStabGroup}. However, other states are stabilized by $4$ elements of $\overline{\mathcal{C}}_2$, but by only $2$ elements in $\barHC$ (see Footnote \ref{C2Footnote}). Accordingly, such states are found in one of the $g_{576}$ graphs in Figure \ref{PhaseConnectedWG288ContractedGraph}. Still other states are stabilized by $4$ elements of $\overline{\mathcal{C}}_2$, but only by the identity in $\barHC$, and reside in the single copy of $g_{1152}$ in Figure \ref{PhaseConnectedWG288ContractedGraph}.

The $\mathcal{C}_2$ reachability graph of $\ket{D^3_1}$ contracts to a $6$-vertex graph, seen to the right of Figure \ref{PhaseConnectedWG288ContractedGraph}, after identifying vertices connected by entropy-preserving circuits. While the contracted graph in Figure \ref{PhaseConnectedWG288ContractedGraph} has $6$ vertices, we only ever observe $5$ different entropy vectors among those vertices. We address this point further in the discussion. The $5$ entropy vectors of the $\ket{D^3_1}$ contracted graph are listed in Table \ref{tab:WStateEntropyVectorTable}.

All remaining Dicke states, those of the form $\ket{D^{n}_k}$ with $1 < k < n-1$, are stabilized by only $2$ elements in $\overline{\mathcal{C}}_2$. For any $\ket{D^{n}_k}$ of this form, its stabilizer subgroup under $\overline{\mathcal{C}}_2$ action is given by
\begin{equation}\label{AllOtherDStabilizer}
\mathcal{S}_{\overline{\mathcal{C}}_2}\left(\ket{D^n_k}\right) = \{\mathbb{1},\, C_{1,2}C_{2,1}C_{1,2}\}, \quad \forall \, 1 < k < n-1.
\end{equation}
We again note that the stabilizer group%
\footnote{The operator $C_{1,2}C_{2,1}C_{1,2}$ is the gate $SWAP_{1,2}$. Dicke states are symmetric under $SWAP_{1,2}$, a feature which is likely interpretable at the group level. We acknowledge a referee for this observation.} %
in Eq.\ \eqref{AllOtherDStabilizer} is also contained completely within $\barHC$, and therefore the left coset space $\barHC / \mathcal{S}_{\overline{\mathcal{C}}_2}\left(\ket{D^n_k}\right)$ consists of $576$ elements. 

The reachability graph for $\ket{D^{n}_k}$ under $\barHC$, which we denote $g_{576}$, has $576$ vertices. The left panel of Figure \ref{G576WithContractedGraph} depicts $g_{576}$, specifically for the state $\ket{D^4_2}$. Reachability graphs with $576$ vertices, under $\barHC$ action, are never observed for stabilizer states. Again, as with $g_{288^*}$, the graph $g_{576}$ corresponds to non-stabilizer states which are non-trivially stabilized by $\overline{\mathcal{C}}_n$.

After identifying vertices in $g_{576}$ connected by entropy-preserving operations, we are left with a contracted graph of $9$ vertices shown on the right of Figure \ref{G576WithContractedGraph}. These $9$ vertices are colored by $6$ different entropy vectors, with maximal coloring beginning at $4$ qubits. Among the $6$ entropy vectors in this contracted graph, there are symmetries shared among cyan, magenta, and yellow vectors, and separately among red, blue, and green vectors. The specific $6$ entropy vectors for the $\ket{D^4_2}$ contracted graph are given in Table \ref{tab:D42EntropyVectors}.

\begin{figure}[h]
        \centering
        \includegraphics[width=15cm]{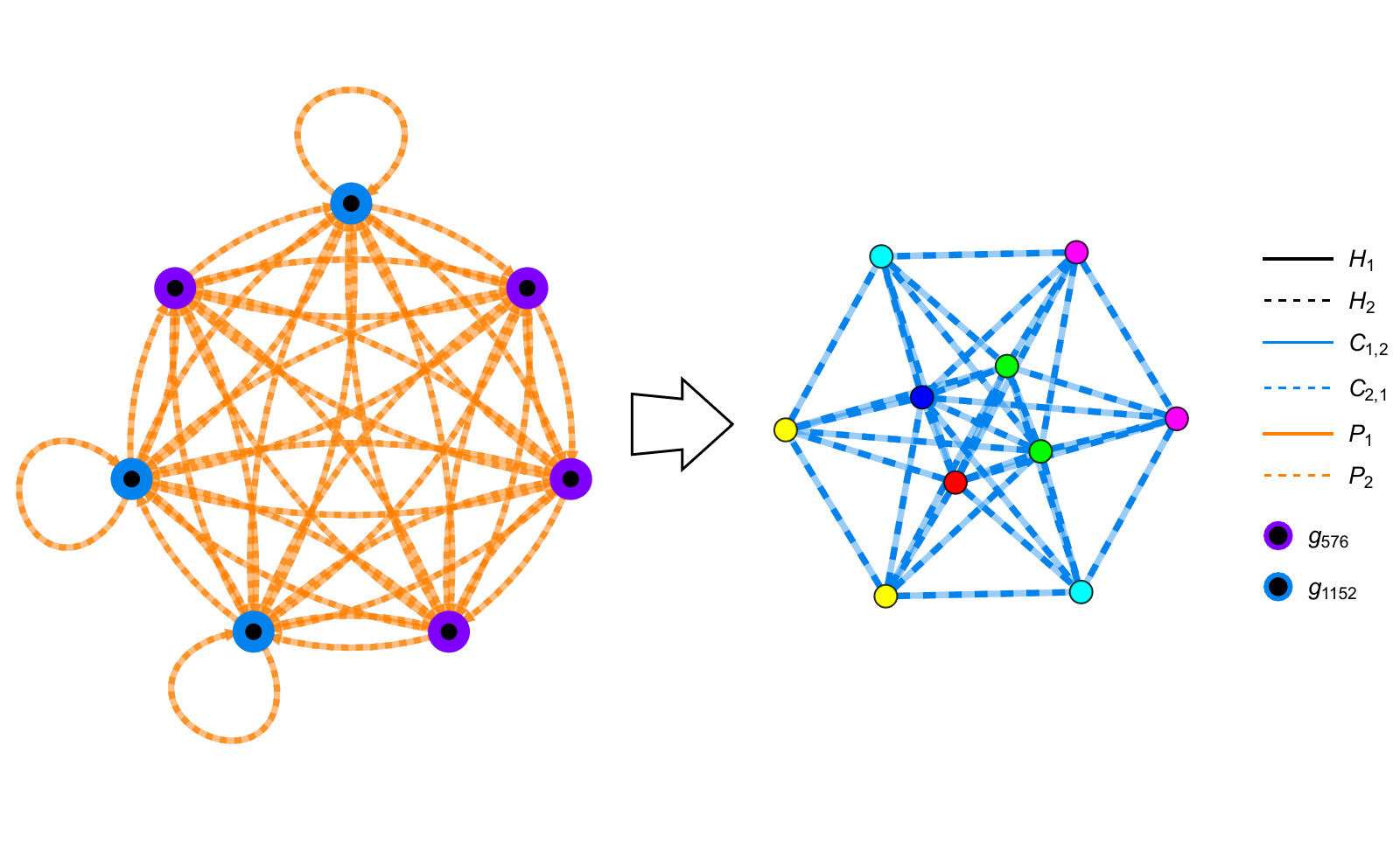}
        \caption{Reachability graph of $\ket{D^4_2}$ under $\overline{\mathcal{C}}_2$ (left), and its associated contracted graph (right). We display the $5760$-vertex reachability graph as a network of reachability $\HC$ graphs $g_{576}$ and $g_{1152}$, connected by $P_1$ and $P_2$ gates. The contracted graph contains $10$ vertices, but we only ever observe $6$ entropy vectors due to how the $g_{576}$ and $g_{1152}$ copies connect under phase action. The $6$ different entropy vectors shown are given in Table \ref{tab:D42EntropyVectors}.}
        \label{PhaseConnectedG576ContractedGraph}
    \end{figure}

Acting with the full group $\overline{\mathcal{C}}_2$ on $\ket{D^{n}_k}$, for $1 < k < n-1$, generates an orbit of $5760$ states. The 
$\mathcal{C}_2$ reachability graph of $\ket{D^{n}_k}$ therefore has $5760$ vertices, and is depicted in the left panel of Figure \ref{PhaseConnectedG576ContractedGraph} for the case of $\ket{D^4_2}$. As before, we depict the full $5760$-vertex reachability graph as $7$ attached copies of different $\HC$ reachability graphs $g_{576}$ and $g_{1152}$.

The $5760$-vertex reachability graph in Figure \ref{PhaseConnectedG576ContractedGraph} consists of $4$ copies of $g_{576}$ and $3$ copies of $g_{1152}$, all connected via $P_1$ and $P_2$ operations. While every state in the full $5760$-vertex reachability graph is stabilized by $2$ elements of $\overline{\mathcal{C}}_2$, some states have a stabilizer group completely contained within $\barHC$. States stabilized by $2$ elements of $\barHC$ are found in one of the $4$ copies of $g_{576}$ in Figure \ref{PhaseConnectedG576ContractedGraph}. Alternatively, states which are stabilized by $2$ elements of $\overline{\mathcal{C}}_2$, but only the identity in $\barHC$, are found in one of the $3$ copies of $g_{1152}$.

If we identify vertices connected by entropy-preserving operations in the $\mathcal{C}_2$ reachability graph of $\ket{D^4_2}$, we are left with a contracted graph containing $10$ vertices shown to the right of Figure \ref{PhaseConnectedG576ContractedGraph}. While this contracted graph has $10$ vertices, we only ever observe $6$ different entropy vectors among those $10$ vertices. We again return to this point in the discussion. The contracted graph in Figure \ref{PhaseConnectedG576ContractedGraph} also reflects the symmetry among magenta, cyan, and yellow vertices observed in Figure \ref{G576WithContractedGraph}. These $6$ entropy vectors which can be generated from $\ket{D^4_2}$ under $\mathcal{C}_2$ are given in Table \ref{tab:D42EntropyVectors}.

In this subsection we extended our analysis beyond the stabilizer states, building contracted graphs for non-stabilizer Dicke states under the action of $\barHC$ and $\overline{\mathcal{C}}_2$. States $\ket{D^n_k}$, for $k \neq n$, are particularly interesting at $n \geq 3$ qubits as they comprise a class of non-stabilizer states that are non-trivially stabilized by elements of $\overline{\mathcal{C}}_n$. We constructed the two possible reachability graphs for $\ket{D^n_k}$, one for states $\ket{D^n_1}$ and $\ket{D^n_{n-1}}$, and the other for all $\ket{D^n_k}$ with $1 < k < n-1$. We described how each Dicke state $\mathcal{C}_2$ reachability graph corresponds to a connection of $\HC$ reachability graphs $g_{288^*},\, g_{576},$ and $g_{1152}$ under $P_1$ and $P_2$ operations.

We built the contracted graphs for each $\ket{D^n_k}$ $\HC$ and $\mathcal{C}_2$ reachability graph. We illustrated that states $\ket{D^n_1}$ and $\ket{D^n_{n-1}}$ can realize $5$ different entropy vectors under $\mathcal{C}_2$. Alternatively, states of the form $\ket{D^n_k}$ with $1 < k < n-1$ can achieve $6$ different entropy vectors under $\mathcal{C}_2$. In the next subsection we completely generalize to an argument for $\mathcal{C}_n$ action on arbitrary entropy vectors. We use our construction up to this point to bound the entropy vector possibilities that can be achieved for any state under $n$-qubit Clifford action.

\subsection{Entanglement in $n$-Qubit Clifford Circuits}

We now use our results to present an upper bound on entropy vector evolution in Clifford circuits, for arbitrary qubit number. We begin by determining the subset of $\mathcal{C}_n$ operations which cannot modify the entanglement entropy of any state. We then build a contracted graph by identifying the vertices in the $\mathcal{C}_n$ Cayley graph that are connected by entropy-preserving circuits.

Local actions, i.e.\ all operations which act only on a single qubit in some $n$-qubit system, will always preserve a state's entropy vector. When considering action by the Clifford group $\mathcal{C}_n$, the subgroup of all local actions is exactly%
\footnote{If a quantum circuit, acting on a finite number of qubits, preserves the entanglement of all quantum states, then it must be expressible as the tensor product of local unitaries acting independently on individual qubits. Moreover, if the circuit is composed only of Clifford gates, it must be expressible as the product of local Clifford gates, i.e. the tensor product of Hadamard and phase gates.} %
the group generated by $n$-qubit Hadamard and phase gates, which we denote $(HP)_n$. We build $(HP)_n$ as the direct product \cite{Keeler:2023xcx}
\begin{equation}
    (HP)_n \equiv \prod_{i=1}^n \langle H_i,\,P_i \rangle.
\end{equation}
Since $(HP)_n$ is a direct product, and $|\overline{\langle H_i,\,P_i \rangle}| = 24$, the order of $|(\overline{HP})_n|$ is just $24^n$. The order of the phase-quotiented $n$-qubit Clifford group is likewise known \cite{Walter2016}. We can compute $|\overline{\mathcal{C}}_n|$ as
\begin{equation}
    |\overline{\mathcal{C}}_n| = 2^{n^2+2n}\prod_{j=1}^n(4^j-1).
\end{equation}

Generating the right coset space $(\overline{HP})_n \backslash \overline{\mathcal{C}}_n$ identifies all elements in $\overline{\mathcal{C}}_n$ equivalent up to local gate operations. Invoking Lagrange's theorem (Eq.\ \eqref{LagrangeTheorem}) allows us to compute the size of $(\overline{HP})_n \backslash \overline{\mathcal{C}}_n$ as
\begin{equation}\label{NonLocalGroupOrder}
    \frac{|\overline{\mathcal{C}}_n|}{|(\overline{HP})_n|} = \frac{2^{n^2-n}}{3^n}\prod_{j=1}^n(4^j-1).
\end{equation}

It is important to note that $(\overline{HP})_n$ is not a normal subgroup of $\overline{\mathcal{C}}_n$, which we can immediately verify by considering any Hadamard operation $H_j \in (\overline{HP})_n$. The element
\begin{equation}
    C_{i,j}H_jC_{i,j}^{-1} \notin \overline{\langle H_i,\,P_i,\,H_j,\,P_j \rangle},
\end{equation}
which violates the necessity that any normal subgroup be invariant under group conjugation. Accordingly, $(\overline{HP})_n$ does not generate a quotient of $\overline{\mathcal{C}}_n$.

The coset space $(\overline{HP})_n \backslash \overline{\mathcal{C}}_n$ partitions $\overline{\mathcal{C}}_n$ into sets of Clifford circuits which are equivalent up to local action. Consequently, Eq.\ \eqref{NonLocalGroupOrder} provides an upper bound on the number of entropy vectors that can possibly be generated under any $n$-qubit Clifford circuit, for any arbitrary quantum state. This upper bound is equivalently captured by directly building a contracted graph from the $\overline{\mathcal{C}}_n$ Cayley graph, and counting the number of vertices. The right panel of Figure \ref{FullC2WithContractedGraph} illustrates the $20$-vertex contracted graph of the $\overline{\mathcal{C}}_2$ Cayley%
\footnote{Formally, the left panel of Figure \ref{FullC2WithContractedGraph} depicts the reachability graph for some set of states, rather than the Cayley graph of $\overline{\mathcal{C}}_2$. However, since the particular class of states is stabilized by only the identity in $\overline{\mathcal{C}}_2$, the reachability graph in the left panel of Figure \ref{FullC2WithContractedGraph} is exactly the phase-modded $\mathcal{C}_2$ Cayley graph.} %
graph. Table \ref{tab:EntropicDiversity} gives the explicit number of entropy vectors that can be achieved using $n \leq 5$ qubit Clifford circuits.
\begin{table}[h]
    \centering
    \begin{tabular}{|c||c|}
    \hline
    $n$ & $|\overline{\mathcal{C}}_n|/|(\overline{HP})_n|$\\
    \hline
    \hline
    $1$ & $1$ \\
    \hline
    $2$ & $20$ \\
    \hline
    $3$ & $6720$ \\
    \hline
    $4$ & $36556800$ \\
    \hline
    $5$ & $3191262412800$ \\
    \hline
    \end{tabular}
\caption{Maximum number of entropy vectors that can be generated using elements of the $n$-qubit Clifford group, for $n \leq 5$.}
\label{tab:EntropicDiversity}
\end{table}

In Eq.\ \eqref{NonLocalGroupOrder} we count the right cosets of $\overline{\mathcal{C}}_n$ by the subgroup of entropy-preserving operations. This upper bound equivalently constrains the number of entropy vectors which can be realized by a generic quantum state, stabilized by only $\mathbb{1} \in \overline{\mathcal{C}}_n$, under any Clifford circuit. However, we can tighten this bound for states which are non-trivially stabilized by some subset of $\overline{\mathcal{C}}_n$. For a state $\ket{\psi}$ with stabilizer group $\mathcal{S}_{\overline{\mathcal{C}}_n}(\ket{\psi})$, the number of achievable entropy vectors is bounded by the size of the double coset space $(\overline{HP})_n \backslash \overline{\mathcal{C}}_n/\mathcal{S}_{\overline{\mathcal{C}}_n}(\ket{\psi})$. As by Eq.\ \eqref{DoubleCosetOrder}, the size of $(\overline{HP})_n \backslash \overline{\mathcal{C}}_n/\mathcal{S}_{\overline{\mathcal{C}}_n}(\ket{\psi})$ is
\begin{equation}\label{GeneralContractedGraphOrder}
    |(\overline{HP})_n \backslash \overline{\mathcal{C}}_n/\mathcal{S}_{\overline{\mathcal{C}}_n}(\ket{\psi})| = \frac{1}{|(\overline{HP})_n||\mathcal{S}_{\overline{\mathcal{C}}_n}(\ket{\psi})|} \sum_{(h,s) \in (\overline{HP})_n \times \mathcal{S}_{\overline{\mathcal{C}}_n}(\ket{\psi})} |\overline{\mathcal{C}}_n^{(h,s)}|, 
\end{equation}
where $\overline{\mathcal{C}}_n^{(h,s)}$ is defined by Eq.\ \eqref{DoubleCosetDefinition}.

Applying Eq.\ \eqref{GeneralContractedGraphOrder} when $\ket{\psi}$ is a stabilizer state dramatically reduces the number of possible entropy vectors that can be reached under $\mathcal{C}_n$. Specifically, when restricting to group action by $\HC$, Eq.\ \eqref{GeneralContractedGraphOrder} computes the vertex count for each of the five contracted graphs shown in Figures \ref{G24WithContractedGraph} -- \ref{FullC2WithContractedGraph}.

In this subsection we provided an upper bound on the number of entropy vectors that can be generated by any Clifford circuit, at arbitrary qubit number. For a generic quantum state, we showed that the number of possible entropy vectors is bounded by the size of the right coset space $(\overline{HP})_n \backslash \overline{\mathcal{C}}_n$. Alternatively, for states stabilized by additional elements in $\overline{\mathcal{C}}_n$, the number of possible entropy vectors is bounded by the size of the double coset space $(\overline{HP})_n \backslash \overline{\mathcal{C}}_n/\mathcal{S}_{\overline{\mathcal{C}}_n}(\ket{\psi})$.

\section{From Entropic Diversity to Holographic Interpretation}\label{sec:diversity}

The contracted graphs in Section \ref{ContractedGraphsSection} illustrate the diversity of entropy vectors on $\HC$ and $\mathcal{C}_2$ reachability graphs. We now analyze this entropic diversity as we move towards a holographic interpretation of our results. We begin by considering the maximum number of different entropy vectors that can be found on each of the $\HC$ and $\mathcal{C}_2$ graphs studied in the above section, as well as the minimum number of qubits needed to realize that maximal diversity. We explore the implications of entropic diversity and graph diameter as constraining the transformations of a geometric gravitational dual in holography. We then present the number of $\HC$ subgraphs, including isomorphic subgraphs with different entropic diversities, as we increase qubit number. We remark how our contracted graphs encode information about entropy vector evolution through entropy space.

\subsection{Clifford Gates in Holography}

The AdS/CFT conjecture \cite{Maldacena:1997re} is a bulk/boundary duality which relates gravitational objects in an asymptotically hyperbolic spacetime, evaluated at some fixed timeslice $\Sigma$, with computable properties of a quantum-mechanical system on the boundary of that spacetime $\partial \Sigma$. For a special class of quantum states known as holographic states, the Ryu-Takayanagi formula relates all components of the state's entropy vector to areas of extremal surfaces in the dual gravity theory \cite{Ryu:2006bv,Faulkner:2013ana}. In this way, a description of the spacetime geometry in $\Sigma$ is inherited from knowledge of the entanglement structure on $\partial \Sigma$. For this relation to hold, holographic states are required to have an entropy vector structure which satisfies a set of holographic entropy inequalities \cite{Bao2015,HernandezCuenca2019}. One holographic inequality, the monogamy of mutual information (MMI) \cite{Hayden2013}, reads
\begin{equation}\label{MMI}
    S_{AB} + S_{AC} + S_{BC} \geq S_{A} + S_{B} + S_{C} + S_{ABC},
\end{equation}
and must be satisfied for all%
\footnote{It is important to note that each $A,B,C \subseteq \partial \Sigma$ may separately correspond to the disjoint union of multiple qubits in the $n$-party boundary theory. Accordingly, the MMI inequality in Eq.\ \eqref{MMI} must hold for disjoint triples $\{A,B,C\}$, as well as those of the form $\{AB,C,DE\}$ or $\{ABC,DE,F\}$, and so on. Furthermore, holographic states must saturate or satisfy MMI for all permutations among any chosen $A,B,C \subseteq \partial \Sigma$.} %
$A,B,C \subseteq \partial \Sigma$. While MMI constitutes only one of many holographic entropy inequalities, it arises at four qubits, while all other holographic inequalities require more parties.

Understanding the entropy-vector dynamics of a state in $\partial \Sigma$ gives insight into bulk geometric transformations in $\Sigma$. When a local operator acts on $\ket{\psi}$ and modifies its entropy vector to another vector within the holographic entropy cone, geodesics in the dual spacetime geometry are likewise modified in accordance with the RT formula. Consequently, analyzing how a group of operators transforms the entropy vector of a state can reveal how gate action on $\partial \Sigma$ alters geometries in $\Sigma$. When a sequence of Clifford gates causes the state to violate holographic inequalities, the geometry may be only a semi-classical approximation.

The distance between vertices on reachability graphs encodes a natural notion of circuit complexity. Entropy vectors which populate the same reachability graph, e.g.\ under $\barHC$ or $\overline{\mathcal{C}}_2$, may be considered close in the sense that a limited number of gate applications is required to transform a state with one entropy vector into some state with another. The gravitational dual geometries of states with ``nearby'' entropy vectors may be considered close in a similar sense, since a small number of manipulations are needed to transform one dual geometry into each other. 

Some $n$-qubit stabilizer states have entropy vectors which violate the holographic entropy inequalities, beginning at $n=4$. Since stabilizer entanglement is generated by bi-local gates, $2$-qubit Clifford operations are sufficient to generate all stabilizer entropy vectors in an $n$-qubit system. We can therefore explore the transition from holographic entropy vectors to non-holographic stabilizer entropy vectors by observing entropy vector evolution under $\mathcal{C}_2$. In the following subsections we discuss how entropic diversity on $\HC$ and $\mathcal{C}_2$ reachability graphs can inform us about states which are geometrically close, and not so close, in the dual gravitational theory.

\subsection{Maximal Entropic Diversity for Stabilizer States}

Each $\HC$ and $\mathcal{C}_2$ reachability graph describes the full orbit of some state $\ket{\psi} \in \Hil$ under the action of $\barHC$ or $\overline{\mathcal{C}}_2$ respectively. While we can construct reachability graphs for an arbitrary $n$-qubit quantum state, including states with arbitrary entanglement structure, the set of possible entropy vectors that can be reached under $\HC$ and $\mathcal{C}_2$ remains bounded at the operator level. For a given reachability graph, we refer to the maximum number of possible entropy vectors that can be generated in that graph as the maximal entropic diversity of the graph.

Table \ref{tab:MaximalHCColoringTable} gives each stabilizer state $\HC$ reachability graph, and the maximal entropic diversity determined by its contracted graph. For certain subgraphs, such as $g_{144},\,g_{288},$ and $g_{1152}$, the number of qubits needed to realize the maximal entropic diversity is higher than the number of qubits at which each graph first appears.
\begin{table}[h]
    \centering
    \begin{tabular}{|c||c|c|c|}
    \hline
    $\HC$ Graph & \multicolumn{1}{|p{2.6cm}|}{\centering Max Entropic \\ Diversity} & \multicolumn{1}{|p{2.7cm}|}{\centering Stab. Qubit\\ Num. Appears} & \multicolumn{1}{|p{3.5cm}|}{\centering Stab. Qubit Num.\\ Max Diversity}\\
    \hline
    \hline
    $g_{24}$ & $2$  & $2$ & $2$ \\
    \hline
    $g_{36}$ & $2$  & $2$ & $2$ \\
    \hline
    $g_{144}$ & $5$  & $3$ & $6$ \\
    \hline
    $g_{288}$ & $5$  & $3$ & $6$ \\
    \hline
    $g_{1152}$ & $18$  & $4$ & $7$ or $8$ \\
    \hline
    \end{tabular}
\caption{Stabilizer state $\HC$ graphs listed alongside their maximal entropic diversities, set by contracted graphs. We give the qubit number when each graph is first observed for stabilizer states, and the minimum qubit number needed to realize the maximal entropic diversity for stabilizer states. We have found $g_{1152}$ graphs with maximal diversity for $8$-qubit stabilizer states, but have not completely ruled out a maximally diverse $g_{1152}$ graph at $7$ qubits since an exhaustive search is computationally difficult.}
\label{tab:MaximalHCColoringTable}
\end{table}

The entropy vectors on $g_{24}$ and $g_{36}$ correspond to maximal and minimal $2$-qubit entanglement, and can therefore be achieved by entangling only $2$ qubits in an $n$-party system. These two entropy vectors are close in the sense that they are connected by a single $C_{1,2}$ action.  Since this single gate acts on only $2$ out of the $n$ qubits, we expect states with these entropy vectors to admit close dual (possibly semi-classical) geometries. Analogously, altering only small segments of the boundary of a holographic state will affect its geometry only inside the entanglement wedge of the union of these segments. 

For larger reachability graphs, the graph diameter upper bounds the $\barHC$ gate distance, and thus the geometric closeness, of the included entropy vectors. In particular, $g_{1152}$ is the $\HC$ reachability graph for generic quantum states, and its maximal entropic diversity gives an upper bound on the number of distinct entropy vectors, and thus the number of distinct semi-classical geometries, reachable under $\HC$ action. 

We additionally compile the entropic diversity data for all stabilizer state $\mathcal{C}_2$ reachability graphs. As shown throughout Section \ref{ContractedGraphsSection}, every $\mathcal{C}_2$ graph is a complex of $\HC$ subgraphs attached by $P_1$ and $P_2$ edges. Table \ref{tab:MaximalC2ColoringTable} lists the different $\mathcal{C}_2$ complexes, and the maximal entropic diversity of each.
\begin{table}[h]
    \centering
    \begin{tabular}{|c|c||c|c|c|}
    \hline
    Fig. & $\mathcal{C}_2$ Graph & \multicolumn{1}{|p{2.6cm}|}{\centering Max Entropic \\ Diversity} & \multicolumn{1}{|p{2.7cm}|}{\centering Stab. Qubit\\ Num. Appears} & \multicolumn{1}{|p{3.5cm}|}{\centering Stab. Qubit Num.\\ Max Diversity}\\
    \hline
    \hline
    \ref{PhaseConnectedG24:G36}, \ref{PhaseConnectedG24:g36ContractedGraph} & $g_{24} + g_{36}$ & $2$  & $2$ & $2$\\
    \hline
    \ref{HhCcPhaseOverlay}, \ref{PhaseConnectedg144_288ContractedGraph} & $3 \cdot g_{144}+g_{288}$ & $5$  & $3$ & $6$\\
    \hline
    \ref{FullC2WithContractedGraph} & $10 \cdot g_{1152}$ & $20$ & $4$ & $7$ or $8$\\
    \hline
    \end{tabular}
\caption{Each stabilizer state $\mathcal{C}_2$ graph, built of attached $\HC$ subgraphs. Each graph is listed alongside its maximal entropic diversity, set by its contracted graph. We give the first time each graph appears as a stabilizer state orbit, and the first time each graph achieves maximal entropic diversity for stabilizer states.}
\label{tab:MaximalC2ColoringTable}
\end{table}

The addition of $P_1$ and $P_2$ enables two more entropy vectors to be reached by states in a $g_{1152}$ subgraph. Although this section has so far concentrated on the stabilizer states, the $10 \cdot g_{1152}$ $\mathcal{C}_2$ complex is actually the generic reachability graph for arbitrary quantum states, which are not stabilized by any non-identity element of a given two-qubit Clifford group. Accordingly, the $20$ entropy vectors in this complex constrain the possible unique entropy vectors that can be generated by starting with a generic quantum state and acting with $2$-qubit Clifford operations.

In this subsection we provided Tables \ref{tab:MaximalHCColoringTable}--\ref{tab:MaximalC2ColoringTable} which detailed the maximal entropic diversity of each stabilizer state $\HC$ and $\mathcal{C}_2$ reachability graph. Additionally, we provided the minimal system size needed to realize each maximal entropic diversity in a stabilizer state orbit. Note that for other quantum states with the same reachability graphs, maximal entropic diversity could be achieved at lower qubit numbers. We speculated that the maximal entropic diversity of reachability graphs constrains the available transformations, and that the graph diameter constrains the dissimilarity, of the dual geometries that can be generated from $\HC$, or $\mathcal{C}_2$, action on the boundary state. In the next subsection we analyze the number, and diversity, of stabilizer state reachability graphs as the number of qubits in the system increases.

\subsection{$\mathcal{C}_2$ Subgraph Count by Qubit Number}

The number of times each stabilizer state $\mathcal{C}_2$ reachability graph in Section \ref{ContractedGraphsSection} occurs in the set of $n$-qubit stabilizer states increases with every qubit added to the system. Furthermore, as we increase qubit number we observe different entropic diversities which are possible on $\mathcal{C}_2$ reachability graphs. Table \ref{tab:SubgraphCountTable} gives a count for each variety of stabilizer state $\mathcal{C}_2$ graph, with increasing qubit number, for $n \leq 5$ qubits.
\begin{table}[h]
    \centering
    \begin{tabular}{|c||c|c|c|}
     \hline
     & \multicolumn{3}{|c|}{$\mathcal{C}_2$ Graph} \\
    \hline
    Qubit \# & $g_{24}/g_{36}$ & $g_{144}/g_{288}$ & $g_{1152}$\\
    \hline
    \hline
    2 & 1 (2) & 0  & 0\\
    \hline
    3 & 6 (2) & 1 (3) & 0\\
    \hline
    4 & 60 (2) & 12 (3), 18 (4) & 1 (2), 9 (4)\\
    \hline
    5 & 1080 (2) & 180 (3), 1080 (4) & 18 (2), 216 (4), 486 (6), 540 (7)\\
    \hline
    \end{tabular}
\caption{Distribution of stabilizer state $\mathcal{C}_2$ reachability graphs, and their different entropic diversities, for $n \leq 5$ qubits. The first number in each cell gives the number of occurrences for each $\mathcal{C}_2$ subgraph, while the number in parentheses gives the entropic diversity of each subgraph variation.}
\label{tab:SubgraphCountTable}
\end{table}

The overall count of each $\mathcal{C}_2$ subgraph increases as the size of the system grows. Graph $g_{1152}$ however, shown in the final column of Table \ref{tab:SubgraphCountTable}, has an occurrence count which increases the fastest with qubit number. As expected, when the system size grows large the percentage of states stabilized by any non-identity $2$-qubit Clifford subgroup decreases. 

Subgraphs $g_{144}/g_{288}$ can have an entropic diversity of $3,\,4,$ or $5$, while states in a $g_{1152}$ $\mathcal{C}_2$ complex can reach up to $20$ different entropy vectors. As qubit number increases the number of entanglement possibilities grows, yielding more complex entropy vectors. Entropy vectors with sufficient complexity will change the maximal number of allowed times under $\mathcal{C}_2$ action. We therefore expect the number of $g_{144}/g_{288}$ graphs with $5$ entropy vectors, and $g_{1152}$ $\mathcal{C}_2$ graphs with $20$ entropy vectors, to dominate the subgraph occurrence count in the large system limit. For larger subgraphs, e.g.\ those composed of $g_{1152}$ subgraphs, understanding the precise distribution of entropic diversity for arbitrary qubit number presents a challenging problem, which we leave for future work. We now conclude this section with a discussion of Dicke state entropic diversity in $\HC$ and $\mathcal{C}_2$ reachability graphs.

\subsection{Maximum Entropic Diversity for Dicke States}

We now analyze the entropic diversity of the Dicke state $\ket{D^n_k}$ reachability graphs in Section \ref{DickeStateSubsection}. Subgraphs $g_{288^*}$ and $g_{576}$ correspond to the two possible $\ket{D^n_k}$ orbits under $\barHC$ action, shown in Figures \ref{WStateG288WithContractedGraph}--\ref{G576WithContractedGraph}. Under the full action of $\overline{\mathcal{C}}_2$, $P_1$ and $P_2$ edges attach copies of $g_{288^*}, g_{576},$ and $g_{1152}$ together, creating the graph complexes seen in Figures \ref{PhaseConnectedWG288ContractedGraph}--\ref{PhaseConnectedG576ContractedGraph}. In Table \ref{tab:MaximalDiversityTableDickeStates} we present the maximal entropic diversity of each $\ket{D^n_k}$ $\HC$ and $\mathcal{C}_2$ reachability graph, as determined by their contracted graphs.
\begin{table}[h]
    \centering
    \begin{tabular}{|c|c||c|c|c|}
    \hline
    Fig. & Graph & \multicolumn{1}{|p{2.4cm}|}{\centering Max Entropic \\ Diversity} & \multicolumn{1}{|p{2.4cm}|}{\centering First Appears \\ for $\ket{D^n_k}$} & \multicolumn{1}{|p{3cm}|}{\centering Max Diversity\\ for $\ket{D^n_k}$}\\
    \hline
    \hline
    \ref{WStateG288WithContractedGraph} & $g_{288^*}$ & $5$  & $3$ & $5$ \\
    \hline
    \ref{PhaseConnectedWG288ContractedGraph} & $2 \cdot g_{288^*} + 2 \cdot g_{576}+ g_{1152}$ & $6$  & $3$ & $5$\\
    \hline
    \ref{G576WithContractedGraph} & $g_{576}$ & $9$ & $4$ & $6$ \\
    \hline
    \ref{PhaseConnectedG576ContractedGraph} & $4 \cdot g_{576} + 3 \cdot g_{1152}$ & $10$ & $4$ & $6$\\
    \hline
    \end{tabular}
\caption{All $\HC$ reachability graphs (rows $1$ and $3$) and $\mathcal{C}_2$ reachability graphs (rows $2$ and $4$) for Dicke states. We give the maximal entropic diversity of each graph, as set by the contracted graph, as well as the first time the graph appears for Dicke states and the largest entropic diversity achieved among $\ket{D^n_k}$ states. For $\mathcal{C}_2$ graphs in particular, we never observe a $\ket{D^n_k}$ orbit that achieves the maximum number of allowed entropy vectors.}
\label{tab:MaximalDiversityTableDickeStates}
\end{table}

Both $\mathcal{C}_2$ reachability graphs in Table \ref{tab:MaximalDiversityTableDickeStates} do not achieve their maximal entropic diversities as orbits of Dicke states. We expect that a state with sufficiently general entanglement structure, which also shares one of these reachability graphs%
\footnote{Recall that the reachability graphs in Table \ref{tab:MaximalDiversityTableDickeStates} are shared by all states with stabilizer group given by Eqs. \eqref{WStateStabGroup} or \eqref{AllOtherDStabilizer}, and are not restricted to $\ket{D^n_k}$ orbits. Since the entropy vector is a state property, the state structure determines entropy vector complexity and therefore how much an entropy vector can change under some group action.}%
, would realize the maximum allowed number of distinct entropy vectors, though we have not shown this explicitly. In Section \ref{sec:discussion} we speculate on the highly symmetric structure of $\ket{D^n_k}$ entropy vectors as a potential cause for the maximal diversity not being achieved in such graphs.

In this section we analyzed the entropic diversity of reachability graphs studied throughout Section \ref{ContractedGraphsSection}. We detailed each reachability graph achieves its maximal entropic diversity, and speculate implications for the geometric interpretations of state entropy vectors in a dual gravity theory. We demonstrated how certain $\HC$ and $\mathcal{C}_2$ subgraphs appear more frequently with increasing qubit number, as well as how different entropic variations of each subgraph are distributed when the system size grows large. We addressed the notable case of Dicke state reachability graphs, which do not achieve their maximal entropic diversity as orbits of $\ket{D^n_k}$. We will now conclude this work with an overview of our results and some ideas for future research.

\section{Discussion and Future Work}\label{sec:discussion}

In this work we presented a procedure for quotienting a reachability graph to a contracted graph, which allowed us to analyze and bound entropy vector evolution under group action on a Hilbert space. We first constructed a reachability graph, built as a quotient of the group Cayley graph \cite{Keeler:2023xcx}, for a family of states defined by their stabilizer subgroup under the chosen group action. As a group-theoretic object, the vertex set of a reachability graph is the left coset space generated by the stabilizer subgroup for the family of states. We then further quotiented this reachability graph by identifying all vertices connected by edges that preserve the entropy vector of a state. This second graph quotient corresponds to the right coset space generated by the subgroup of elements which leave an entropy vector invariant. The resultant object, after both graph quotients, is a contracted graph. This contracted graph represents the double coset space built of group elements which simultaneously stabilize a family of states, and do not modify an entropy vector.

A contracted graph encodes the evolution of a state entropy vector under group action. Specifically, the number of vertices in a contracted graph strictly bounds the maximal number of distinct entropy vectors that can be found on a reachability graph. The edges of a contracted graph detail the possible changes an entropy vector can undergo through circuits composed of the group generating set. We built $\HC$ and $\mathcal{C}_2$ contracted graphs for all stabilizer states, and demonstrated how the vertex count of each explains the reachability graph entropy distributions observed in our previous work \cite{Keeler2022,Keeler:2023xcx}.

Although we did derive a general upper bound on the number of different entropy vectors that can be reached using any $n$-qubit Clifford circuit starting from an arbitrary quantum state, much of our work focused on $\mathcal{C}_2$ contracted graphs.  However, we could use the same techniques to extend our analysis to $\mathcal{C}_n$, for $n \geq 3$, increasing our generating gate set for additional qubits. In fact, a presentation for $\mathcal{C}_n$ is proposed in \cite{Selinger2013}, using Clifford relations up through $3$ qubits. Understanding precisely how contracted graphs scale with qubit number might offer tighter constraints on achievable entropy vectors in $\mathcal{C}_n$ circuits, and enable us to study more general entropy vector transformations. In AdS/CFT, we only expect systems with arbitrarily large numbers of qubits to be dual to smooth classical qubits.  Consequently, an improved understanding of large-qubit-number contracted graph behavior would strengthen the connection to previous holographic entropy cone work, and could even yield insights for spacetime reconstruction efforts.

While our work in this paper has focused on Clifford circuits, the contracted graph protocol can be applied equally to circuits composed of alternative gate sets (for example, generators of crystal-like subgroups of $SU(n)$ such as $\mathbb{BT}$ \cite{Gustafson:2022xdt}). Prior efforts have explored entanglement diversity \cite{bataille2019quantumcircuitsc} under the action of circuit groups, using various algebraic and geometric arguments. When the chosen gate set generates a finite group of operators, the associated Cayley graph will be finite, as will any graph quotients. For all such cases, a contracted graph analysis follows exactly as in Section \ref{ContractedGraphsSection}, and can be used accordingly to bound entropy vector evolution in different circuit architectures \cite{Negrin:2024tyj}. By exploring different circuit constructions, we can precisely tune our analysis to focus on operations which may be preferred for specific experiments, e.g. arbitrary rotation gate circuits, constructions which replace multiple CNOT gates with Toffoli gates, and architectures that deliberately avoid gates which are noisy to implement.

Alternatively, if the chosen gate set is finite, but generates an infinite set of operators, we can impose a cutoff at some arbitrary fixed circuit depth. This cutoff truncates the associated Cayley graph, and enables an extension of our methods toward the study of universal quantum circuits up to finite circuit depth. Even without an imposed cutoff, we could use our graph analysis to establish bounds on the rate of entanglement entropy per gate application. This description is reminiscent of the notion of entanglement ``velocity'' in universal quantum circuits \cite{Couch:2019zni,Munizzi_2022}.

Although we were originally interested in entropy vector evolution under some chosen gate set, our techniques are sufficiently general to study the evolution of any state property (see footnote \ref{fn:state_property}). Of immediate interest, for example, is the amount of distillable quantum magic present in a state \cite{Bravyi2004,Bao:2022mkc,Leone:2021rzd,Cao:2024nrx,Munizzi:2024huw}, and how this particular measure of non-stabilizerness changes throughout a quantum circuit. Since magic is preserved up to Clifford group action, one subgroup which leaves the amount of magic in a state invariant is exactly the set $\mathcal{C}_n$. 

Given the algebraic nature of contracted graphs, it would be interesting to consider possible algebraic invariants arising from this construction. Recent attempts have been made \cite{klyachko2002coherentstatesentanglementgeometric} to classify entanglement using such algebraic invariants. For example, to each reduced density matrix we can associate a characteristic polynomial, invariant under local action, which yields the components of each entropy vector. We could further consider%
\footnote{We gratefully acknowledge the referee who brought these ideas to our attention.} %
the evolution of local algebraic properties, such as the Gelfand-Kapranov-Zelevinsky invariant \cite{miyake2002multipartiteentanglementhyperdeterminants}, under a group of operators.

In Section \ref{sec:diversity}, we analyzed the maximal entropic diversity of reachability graphs. A reachability graph has maximal entropic diversity when it realizes the maximum number of possible entropy vectors permitted by its contracted graph. We analyzed at which qubit number each $\HC$ and $\mathcal{C}_2$ reachability graph achieves maximal entropic diversity for stabilizer states, and remarked on the growth of entropic diversity with increasing qubit number. 

Since contracted graphs are defined at the operator level, we are also able to extend our analysis to non-stabilizer states. In this paper, we generated all $\HC$ and $\mathcal{C}_2$ contracted graphs for $n$-qubit Dicke states, a class of non-stabilizer states heavily utilized in optimization algorithms \cite{Cerezo:2020jpv,Niroula:2022wvn}. For these states, we derived an upper bound on the number of different entropy vectors that can exist in Dicke state $\HC$ and $\mathcal{C}_2$ reachability graphs. Interestingly, we have not observed $\mathcal{C}_2$ graphs achieve a maximal entropic diversity for Dicke states (see Figures \ref{PhaseConnectedWG288ContractedGraph}--\ref{PhaseConnectedG576ContractedGraph}). The contracted graphs of $g_{288^*}$ and $g_{576}$ permit $6$ and $10$ unique entropy vectors respectively, but we have only ever witnessed $5$ and $9$ entropy vectors for Dicke states with these graphs. We suspect the reason no Dicke state orbit attains its permitted maximal entropy diversity is due to additional $\mathcal{C}_2$ elements which stabilize specifically the highly symmetric entropy vectors of Dicke states \cite{Schnitzer:2022exe,Munizzi:2023ihc}.

In the body of this work, we connected our analysis of entropic diversity to the holographic framework, where entropy vectors admit a description as geometric objects in a dual gravity theory. We used our entropic diversity results to speculate about constraints on geometric transformations in the dual gravity theory, for states which are holographic or near-holographic. We interpret a contracted graph as a coarse-grained map of an entropy vector's trajectory, through entropy space, under a set of quantum gates. Thus, contracted graphs provide information about moving in entropy space, and thereby moving between different entropy cones. 

In future work, we plan to study precisely which Clifford operations move a holographic entropy vector out of, and back into, the holographic entropy cone. Furthermore, we will explore Clifford circuits that transition a stabilizer entropy vector from satisfying, to saturating, to failing holographic entropy conditions, particularly including the monogamy of mutual information (MMI).  We plan to concentrate on MMI since every explicit stabilizer state we have checked either satisfies all holographic inequalities, or violates at least one MMI condition. While \textit{a priori} we have no reason to expect that all stabilizer states which are not holographic necessarily violate MMI in particular, in practice we observe this to be the case empirically for $n \leq 6$ qubits.

\acknowledgments

The authors thank ChunJun Cao, Zohreh Davoudi, Temple He, Sergio Hernandez-Cuenca, Bharath Sambasivam, Howard Schnitzer, Aaron Szasz, and Claire Zukowski for helpful discussions.  CAK and WM are supported by the U.S. Department of Energy under grant number DE-SC0019470 and by the Heising-Simons Foundation ``Observational Signatures of Quantum Gravity'' collaboration grant 2021-2818. JP is supported by the Simons Foundation through \textit{It from Qubit: Simons Collaboration on Quantum Fields, Gravity, and Information}.

\newpage

\begin{appendix}
\addtocontents{toc}{\protect\setcounter{tocdepth}{1}}

\section{Tables of Entropy Vectors}\label{EntropyVectorTables}

Below we include sets of entropy vectors referenced throughout the paper. The states used to generate each entropy vector set are likewise given in bit-address notation. A bit-address is the ordered set of coefficients multiplying each basis ket of an n-qubit system, e.g. the bit-address $(1,0,0,1,0,0,i,i)$ indicates the state $\ket{000}+\ket{011}+i\ket{110}+i\ket{111}$. We order index qubits within each ket from right to left, i.e. the rightmost digit corresponds to the first qubit of the system, while the leftmost digit represents the $n^{\text{th}}$ qubit of an $n$-qubit system.

\subsection{Entropy Vectors for $6$-Qubit Stabilizer Graphs}

Reachability graphs $g_{144}$ and $g_{288}$, shown in Figures \ref{G144WithContractedGraph}--\ref{PhaseConnectedg144_288ContractedGraph}, can be generated by the action of $\barHC$ or $\overline{\mathcal{C}}_2$ on the $6$-qubit state in Eq. \eqref{SixQubit144State}.
\begin{equation}\label{SixQubit144State}
\begin{split}
\frac{1}{8}(1, &-1, 1, 1, -1, 1, 1, 1, 1, -1, 1, 1, 1, -1, -1, -1, 1, -1, -1, -1, -1, 1, -1, -1, -1, 1, 1, \\
& 1, -1, 1, -1, -1, -1, 1, 1, 1, 1, -1, 1, 1, -1, 1, 1, 1, -1, 1, 1, -1, -1, -1, 1,-1, 1, 1, 1,\\
& \qquad  1, -1, 1, -1, -1, -1, 1, 1, 1)\\
\end{split}
\end{equation}

There are $5$ distinct entropy vectors that can be reached in the orbit of Eq. \eqref{SixQubit144State} under $\barHC$ and $\overline{\mathcal{C}}_2$, given in Table \ref{tab:g144g288EntropyVectorTable}. The colors in the table correspond to the vertex colors in Figures \ref{G144WithContractedGraph}--\ref{PhaseConnectedg144_288ContractedGraph}.
\begin{table}[h]
    \centering
    \begin{tabular}{|c||c|}
    \hline
    Label & Entropy Vector\\
    \hline
    \hline
    \fcolorbox{black}{red}{\rule{0pt}{6pt}\rule{6pt}{0pt}} & $\left(1,0,1,1,1,1,1,2,2,2,2,1,1,1,1,2,2,2,2,2,2,2,2,2,2,2,2,2,2,2,2 \right)$\\
    \hline
    \fcolorbox{black}{blue}{\rule{0pt}{6pt}\rule{6pt}{0pt}} & $\left(0,1,1,1,1,1,1,1,1,1,1,2,2,2,2,2,2,2,2,2,2,2,2,2,2,2,2,2,2,2,2 \right)$\\
    \hline
    \fcolorbox{black}{green}{\rule{0pt}{6pt}\rule{6pt}{0pt}} & $\left(1,1,1,1,1,1,1,2,2,2,2,2,2,2,2,2,2,2,2,2,2,2,2,2,2,2,3,3,3,3,2\right)$\\
    \hline
    \fcolorbox{black}{yellow}{\rule{0pt}{6pt}\rule{6pt}{0pt}} & $\left(1,1,1,1,1,1,1,2,2,2,2,2,2,2,2,2,2,2,2,2,2,2,2,2,2,3,3,2,2,3,3 \right)$\\
    \hline
    \fcolorbox{black}{magenta}{\rule{0pt}{6pt}\rule{6pt}{0pt}} & $\left(1,1,1,1,1,1,1,2,2,2,2,2,2,2,2,2,2,2,2,2,2,2,2,2,2,3,2,3,3,2,3\right)$\\
    \hline
    \end{tabular}
\caption{Table of the $5$ entropy vectors found on $g_{144}$ and $g_{288}$ reachability graphs in Figures \ref{G144WithContractedGraph}--\ref{PhaseConnectedg144_288ContractedGraph}. Colors in the leftmost column correspond to the vertex colors of these figures.}
\label{tab:g144g288EntropyVectorTable}
\end{table}

\subsection{Entropy Vectors for $8$-Qubit $g_{1152}$}

To construct the reachability graphs shown in Figure \ref{G1152WithContractedGraph}--\ref{FullC2WithContractedGraph}, we consider the orbit of the $8$-qubit state in Eq. \eqref{EightQubitState} under the action of $\barHC$ and $\overline{\mathcal{C}}_2$.
\begin{equation}\label{EightQubitState}
\begin{split}
\frac{1}{\sqrt{32}}(0, &0, 0, 0, 0, 0, 0, 0, 0, 0, 0, 0, 0, 0, 0, 0, 0, 0, 0, 0, 0, 0, 0, 0, 0, 0, 0,0, 0, 0, 0, 0,0, 0, 0, 0, \\
& \quad 0, 1, 0, -i, 0, -1, 0, -i, 0, 0, 0, 0, 0, 0, 0, 0, i, 0, -1, 0, -i, 0,-1, 0, 0, 0, 0,0, 0, 0,\\
& \quad \quad   0, 0,  0, 0, 0, 0, 0, 0, 0, 0, 0, 0, 0, 0,0, 0, 0, 0, 0,0, 0, 0, 0, 0, 0, 0, 0, 0, 0, 0, 1, 0, -i, 0,  \\
& \quad \quad  0, 0, 0, 0, 0, 0, 0,0, -1, 0, -i, 0,0, i, 0, -1, 0,0, 0, 0, 0, 0,0, 0, 0, -i, 0,-1, 0, 0, 0,   \\
& \quad \quad \quad  0, 0,   0, 0, 0, 0, 0,0, 0, 0, 0, 0,0, 0, 0, 0, 0, 0, 0, 0, 0, 0,0, 0, 0, 0, 0, 0, 0, 0, -i, 0,\\
& \quad \quad \quad \quad  -1, 0, 0,0, 0, 0, 0, 0,0, 0, i, 0, -1,-1, 0, -i, 0, 0, 0, 0, 0, 0, 0, 0, 0, 1, 0, -i, 0,  \\
& \quad \quad \quad \quad 0, 0, 0, 0, 0,0, 0, 0, 0, 0, 0, 0, 0, 0, 0,0, 0, 0, 0, 0, 0, 0, 0, 0, 0, 0, 0,0, 0, 0, 0, 0, \\
& \quad \quad \quad \quad 0, 0, 0, 0, i,0,  1, 0, -i, 0,1, 0, 0, 0, 0, 0, 0,0, 0, 0, 0, 1,0, i, 0, -1, 0,i, 0, 0, 0, 0) \\
\end{split}
\end{equation}

\begin{figure}[h]
        \centering
        \includegraphics[width=\textwidth]{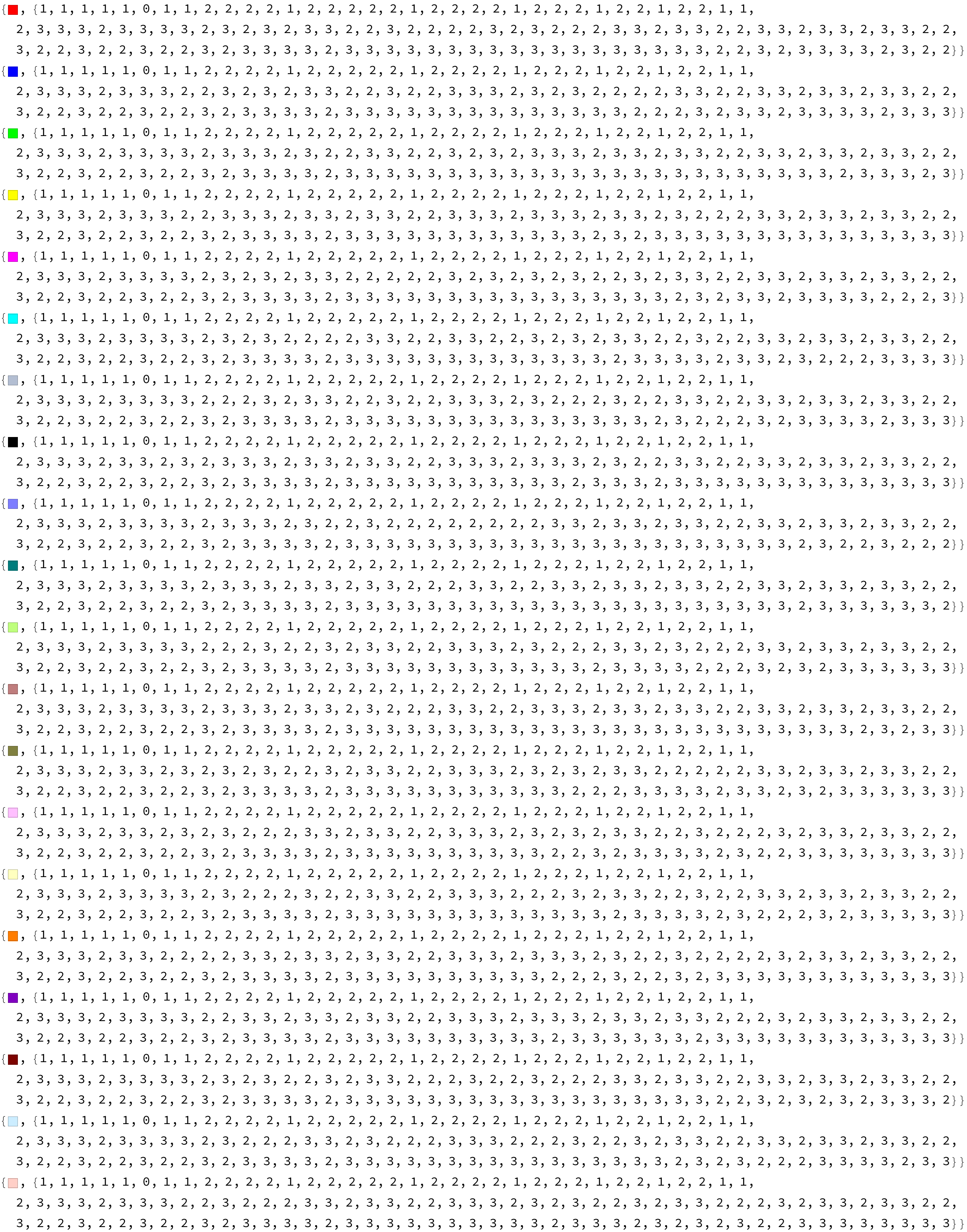}
        \caption{All $8$-qubit entropy vectors reached in the orbit of Eq. \ref{EightQubitState} under the action of $\overline{\mathcal{C}}_2$. Of these $20$ entropy vectors, $18$ can be generated with $\HC$ alone.}
        \label{EightQubitEntropyVectors}
    \end{figure}

The entropy vectors generated along the $\barHC$ and $\overline{\mathcal{C}}_2$ orbits of Eq. \eqref{EightQubitState} are given in Figure \ref{EightQubitEntropyVectors}. The color preceding each entropy vector corresponds to the vertex coloring in Figures \ref{G1152WithContractedGraph}--\ref{FullC2WithContractedGraph}.

\subsection{Entropy Vectors for W-State and Dicke States}

The orbit of $\ket{D^3_1}$ under $\barHC$ and $\overline{\mathcal{C}}_2$ reaches $5$ entropy vectors, built of $4$ different entanglement entropy values. We define these $4$ unique entropy values in Eq.\ \eqref{Entropies31}.
\begin{equation}\label{Entropies31}
    \begin{split}
        s_0&\equiv 1,\\
        s_1&\equiv \frac{2}{3}\log_2\left[\frac{3}{2}\right] +\frac{1}{3}\log_2\left[3\right],\\
        s_2&\equiv \frac{5}{6}\log_2\left[\frac{6}{5}\right] +\frac{1}{6}\log_2\left[6\right],\\
        s_3&\equiv \frac{3-\sqrt{5}}{6}\log_2\left[\frac{6}{3-\sqrt{5}}\right] +\frac{3+\sqrt{5}}{6}\log_2\left[\frac{6}{3+\sqrt{5}}\right],
    \end{split}
\end{equation}

The specific entropy vectors encountered in the $\barHC$ and $\overline{\mathcal{C}}_2$ orbit of $\ket{D^3_1}$ are given in Table \ref{tab:WStateEntropyVectorTable}. Each entropy vector is built from the entanglement entropies given in Eq.\ \ref{Entropies31}. Numerical approximations for each entropy vector were provided in Figure \ref{WStateG288WithContractedGraph} when each first appeared.

\begin{table}[h]
    \centering
    \begin{tabular}{|c||c|}
    \hline
    Label & Entropy Vector\\
    \hline
    \hline
    \fcolorbox{black}{red}{\rule{0pt}{6pt}\rule{6pt}{0pt}} & $(s_1,\,s_1,\,s_1)$\\
    \hline
    \fcolorbox{black}{blue}{\rule{0pt}{6pt}\rule{6pt}{0pt}} & $(s_3,\,s_1,\,s_1)$\\
    \hline
    \fcolorbox{black}{green}{\rule{0pt}{6pt}\rule{6pt}{0pt}} & $(s_1,\,s_3,\,s_1)$\\
    \hline
    \fcolorbox{black}{yellow}{\rule{0pt}{6pt}\rule{6pt}{0pt}} & $(s_0,\,s_0,\,s_1)$\\
    \hline
    \fcolorbox{black}{magenta}{\rule{0pt}{6pt}\rule{6pt}{0pt}} & $(s_2,\,s_2,\,s_1)$\\
    \hline
    \end{tabular}
\caption{Table showing the $5$ entropy vectors seen in Figures \ref{WStateG288WithContractedGraph} and \ref{PhaseConnectedWG288ContractedGraph}, reached in the orbit of $\ket{D^3_1}$ under $\barHC$ and $\overline{\mathcal{C}}_2$. For clarity, we introduce variables in Eq.\ \eqref{Entropies31} to succinctly present each entropy vector.}
\label{tab:WStateEntropyVectorTable}
\end{table}

Similarly for the orbit of $\ket{D^4_2}$ under $\barHC$ and $\overline{\mathcal{C}}_2$, we observe $6$ different entropy vectors. Following the notation of \cite{Munizzi:2023ihc}, we give these $6$ entropy vectors in terms of their $5$ distinct entanglement entropy components, which we list in Eq. \eqref{Entropies42}.
\begin{equation}\label{Entropies42}
    \begin{split}
        s_0&\equiv \frac{5}{6}\log_2\left[\frac{12}{5}\right] +\frac{1}{6}\log_2\left[12\right],\\
        s_1&\equiv \frac{3-\sqrt{5}}{6}\log_2\left[\frac{12}{3-\sqrt{5}}\right] +\frac{3+\sqrt{5}}{6}\log_2\left[\frac{12}{3+\sqrt{5}}\right],\\
        s_2&\equiv \frac{2}{3}\log_2\left[\frac{3}{2}\right] +\frac{1}{3}\log_2\left[6\right],\\
        s_3&\equiv \frac{3-2\sqrt{2}}{6}\log_2\left[\frac{12}{3-2\sqrt{2}}\right] +\frac{3+2\sqrt{2}}{6}\log_2\left[\frac{12}{3+2\sqrt{2}}\right],\\
        s_4&\equiv 1,\\
        s_5&\equiv \frac{2}{3}\log_2\left[\frac{3}{2}\right] +\frac{1}{3}\log_2\left[3\right],\\
        s_6&\equiv \frac{5}{6}\log_2\left[\frac{6}{5}\right] +\frac{1}{6}\log_2\left[6\right].
    \end{split}
\end{equation}

The $5$ entropies in Eq. \eqref{Entropies42} build the $6$ entropy vectors in Table \ref{tab:D42EntropyVectors}.
\begin{table}[h]
    \centering
    \begin{tabular}{|c||c|}
    \hline
    Label & Entropy Vector\\
    \hline
    \hline
    \fcolorbox{black}{red}{\rule{0pt}{6pt}\rule{6pt}{0pt}} & $\left(s_4,\,s_4,\,s_4,\,s_4,\,s_2,\,s_2,\,s_2 \right)$\\
    \hline
    \fcolorbox{black}{blue}{\rule{0pt}{6pt}\rule{6pt}{0pt}} & $\left(s_6,\,s_5,\,s_4,\,s_4,\,s_2,\,s_1,\,s_1 \right)$\\
    \hline
    \fcolorbox{black}{green}{\rule{0pt}{6pt}\rule{6pt}{0pt}} & $\left(s_5,\,s_6,\,s_4,\,s_4,\,s_2,\,s_1,\,s_1 \right)$\\
    \hline
    \fcolorbox{black}{yellow}{\rule{0pt}{6pt}\rule{6pt}{0pt}} & $\left(s_4,\,s_4,\,s_4,\,s_4,\,s_2,\,s_0,\,s_0 \right)$\\
    \hline
    \fcolorbox{black}{magenta}{\rule{0pt}{6pt}\rule{6pt}{0pt}} & $\left(s_4,\,s_4,\,s_4,\,s_4,\,s_2,\,s_2,\,s_2 \right)$\\
    \hline
    \fcolorbox{black}{cyan}{\rule{0pt}{6pt}\rule{6pt}{0pt}} & $\left(s_6,\,s_6,\,s_4,\,s_4,\,s_2,\,s_3,\,s_3 \right)$\\
    \hline
    \end{tabular}
\caption{The $6$ entropy vectors generated by $\HC$ and $\mathcal{C}_2$ on $\ket{D^4_2}$. The vectors appears in Figures \ref{G576WithContractedGraph} and \ref{PhaseConnectedG576ContractedGraph}, and are built using the variables in Eq.\ \eqref{Entropies42}.}
\label{tab:D42EntropyVectors}
\end{table}

\newpage

\end{appendix}

\bibliographystyle{bibliost}
\bibliography{ContractedGraphs} 
\end{document}